\documentclass[5p]{elsarticle}

\usepackage[dvipsnames]{xcolor}
\usepackage[T1]{fontenc}
\usepackage{lmodern}
\DeclareFontFamily{OMX}{lmex}{}
\DeclareFontShape{OMX}{lmex}{m}{n}{<-> lmex10}{}

\usepackage{lipsum}
\usepackage{lineno,subcaption,hyphenat}

\usepackage[hyphens]{url}
\usepackage[breaklinks]{hyperref}

\usepackage{caption}
\usepackage{amsmath, amssymb, amsthm, amsfonts, bm, mathtools}
\usepackage{bigints}
\usepackage{siunitx}
\usepackage{microtype}
\usepackage{wrapfig}
\usepackage{xspace}
\usepackage{placeins}
\usepackage{makecell}
\usepackage{algorithm}
\usepackage[noend]{algpseudocode}
\usepackage{tikz} 
\usetikzlibrary{shapes.geometric,arrows,arrows.meta} 
\usetikzlibrary{shapes.multipart}
\usetikzlibrary{positioning}

\usepackage[numbers]{natbib}
\usepackage{wrapfig}

\usepackage{graphicx}

\DeclareMathAlphabet{\altmathcal}{OMS}{cmsy}{m}{n}

\modulolinenumbers[5]
\makeatletter
\def\BState{\State\hskip-\ALG@thistlm}
\makeatother
\usepackage{turnstile}

\newtheorem*{lemma}{Lemma}

\newcommand\myshade{85}
\colorlet{mylinkcolor}{BrickRed}
\colorlet{mycitecolor}{NavyBlue}
\colorlet{myurlcolor}{Aquamarine}
\hypersetup{
  linkcolor  = mylinkcolor!\myshade!black,
  citecolor  = mycitecolor!\myshade!black,
  urlcolor   = myurlcolor!\myshade!black,
  colorlinks = true,
}

\newcommand{\NGodel}[1]{\ulcorner #1 \urcorner}

\definecolor{lightgray}{gray}{0.94}
\newcommand{\boxsection}[1]{%
  \noindent
  \colorbox{lightgray}{%
\begin{minipage}{\columnwidth}
    #1
    \end{minipage}%
  }%
}

 \newcommand{\hquad}{\hspace{0.05em}} 
 
\usepackage{ctable}

\begin{document}

\begin{frontmatter}

\title{Biological arrow of time: Emergence of  tangled information hierarchies and self-modelling dynamics}

\author{Mikhail~Prokopenko$^{1,2,*}$}
\author{Paul~C.~W.~Davies$^{3}$}
\author{Michael~Harr\'e$^{1,2}$}
\author{Marcus~Heisler$^{1,4}$}
\author{Zdenka~Kuncic$^{1,5,6}$}
\author{Geraint~F.~Lewis$^{5}$}
\author{Ori~Livson$^{1,2}$}
\author{Joseph~T.~Lizier$^{1,2}$}
\author{Fernando~E.~Rosas$^{7,8,9,10}$}

\address{$^{1}$ \hquad The Centre for Complex Systems,  University of Sydney, Sydney, NSW 2006, Australia\\ 
$^{2}$ \hquad School of Computer Science, Faculty of Engineering, University of Sydney, Sydney, NSW 2006, Australia\\
$^{3}$ \hquad The Beyond Center for Fundamental Concepts in Science, Arizona State University, Tempe, AZ, 85287–0506, USA\\
$^{4}$ \hquad School of Life and Environmental Sciences, Faculty of Science, University of Sydney, Sydney, NSW 2006, Australia\\
$^{5}$ \hquad School of Physics, Faculty of Science, University of Sydney, Sydney, NSW 2006, Australia\\
$^{6}$ \hquad The Charles Perkins Centre, University of Sydney, Sydney, NSW 2006, Australia\\
$^{7}$ \hquad Sussex AI and Centre for Consciousness Science, Department of Informatics, University of Sussex, Brighton, BN19RH, UK\\
$^{8}$ \hquad Centre for Psychedelic Research, Department of Brain Science, Imperial College London, London, SW72AZ, UK\\
$^{9}$ \hquad Centre for Complexity Science, Imperial College London, London, SW72AZ, UK\\
$^{10}$ \hquad Center for Eudaimonia and Human Flourishing, University of Oxford, Oxford, OX39BX, UK\\
$^{*}$ \hquad Corresponding author: mikhail.prokopenko@sydney.edu.au
}



\begin{abstract}
We study open-ended evolution by focusing on computational and information-processing dynamics underlying major evolutionary transitions. In doing so, we consider biological organisms as hierarchical dynamical systems that generate regularities in their phase-spaces through interactions with their environment. These emergent information patterns can then be encoded  within the organism's components, leading to self-modelling ``tangled hierarchies''. Our main conjecture is that when macro-scale patterns are encoded within micro-scale components, it creates fundamental tensions (computational inconsistencies) between what is encodable at a particular evolutionary stage and what is potentially realisable in the environment. A resolution of these tensions triggers an evolutionary transition which expands the problem-space, at the cost of generating new tensions in the expanded space, in a continual process. 
We argue that biological complexification can be interpreted computation-theoretically, within the G\"odel--Turing--Post recursion-theoretic framework, as open-ended generation of computational novelty. In general, this process can be viewed as a meta-simulation performed by higher-order systems that successively simulate the computation carried out by lower-order systems. This computation-theoretic argument provides a basis for hypothesising the biological arrow of time.
\end{abstract}


\end{frontmatter}


\tableofcontents

\section{Introduction}


One of the longstanding challenges in biology is the formulation of fundamental principles that can explain why biological complexity is seemingly increasing over time~\cite{arthur1993evolution,lineweaver2013complexity}. 
At the microscopic level, intricate networks of molecular interactions and cellular processes  evolve and adapt over time~\cite{mattick2023rna}. At the mesoscopic level, organisms  appear to continually generate novel forms, behaviours, learning strategies and organisational structures~\cite{mcmillen2024collective}. And at the macroscopic level, ecosystems develop in ways that support complex interactions among species, leading to the emergence of new niches and ecological relationships that, in turn, can drive further evolutionary changes~\cite{dawkins2016extended}. 

The interplay of genetic, environmental and ecological
factors generates never-ending biological innovation and
rich diversity. In general, this seemingly unbounded increase in complexity that characterises evolution on multiple scales has been described as  \textit{open-ended evolution}~\cite{corominas2018zipf}. 
\citet{goldenfeld_life_2011} highlighted a key question which makes understanding the open-ended biological complexity so difficult:
\begin{quote}
``These rules themselves need to evolve, but how? We need an additional set of rules describing the evolution of the original rules. But this upper level of rules itself needs to evolve. Thus, we end up with an infinite hierarchy, an inevitable reflection of the fact that the dynamic we are seeking is inherently self-referential.''
\end{quote}

\noindent
Several approaches to this challenge have been proposed~\cite{goldenfeld_life_2011,pross2011toward,kauffman2014prolegomenon,wong2023roles,heng2023karyotype,sharma2023assembly,wolfram2024why} (see Section~\ref{discussion}). However, developing a clear and comprehensive interpretation of how self-referential dynamics persist and perpetuate across multiple scales remains elusive, as various proposals are scattered across disciplines and lack a unified formal framework.

In this study we address this problem by exploring computational and information-processing dynamics underlying biological complexification.  Our aim is to consider open-ended evolution as a dynamic computational process which successively expands its own problem-space.  
In doing so we interpret major evolutionary transitions (origin of life, formation of eukaryotic cells, emergence of multicellularity, etc.)~\cite{szathmary1995major,west2015major} through the lens of computation theory, aiming to provide a unifying intuitive perspective and improve inter-disciplinary communications.

The core of our argument is as follows.  We consider biological organisms as dynamical systems which may generate regularities in their phase-spaces through their interactions within the environment. In an abstract sense, these emergent regularities can be interpreted as (higher-level) information patterns which may influence the (lower-level) organisms via downward causation
These loops of causation between higher and lower levels are known as tangled hierarchies~\cite{hofstadter_go_1980}.
We hypothesise that these tangled hierarchies can nurture self-modelling capabilities which improve the efficiency of organisms' replication. In other words, self-modelling would allow the organisms to capture compressed representations of the emergent information patterns, by utilising a suitable encoding. However, once such an encoding is adopted, the tangled hierarchies enhanced with self-modelling inevitably generate tensions (inconsistencies) between what is encodable within the current setup and what is possible, that is, realisable in the current environment. Informally, our main argument is that an evolutionary transition resolves these tensions by expanding the problem-space, i.e., by generating a new way to encode extended information patterns.

We begin by reviewing several perspectives on major evolutionary transitions (Section~\ref{bio-view}). This is followed by  a computa\-tion-theoretic argument for the increasing complexity and open-ended evolution developed within the G\"odel--Turing--Post recursion-theoretic framework. This framework formalises the construction of extensible computational systems, such as Turing $\alpha$-oracle machines, and ordinal or recursively generated logics. We argue that this continual process can be interpreted as open-ended meta-simulation which constructs new problem-spaces by resolving computational inconsistencies (Section~\ref{comp-view}). We conclude presenting our background with reviewing  cross-disciplinary insights developed in dynamical systems theory (Section~\ref{dynamical}), as well as complex systems, systems biology, artificial life and machine learning (Section~\ref{cross-disc}). 

Having examined the background studies, we propose a distinction between two types of tangled information hierarchies: with and without self-modelling capability (Section~\ref{tangled}). This distinction allows us to draw a parallel between the open-ended meta-simulation which creates computational novelty and the continual evolutionary process which discovers new phase-spaces along evolutionary transitions in individuality (Section~\ref{self-modelling}).  We then clarify the role of self-reference and fundamental undecidability in forming \textit{the biological arrow of time} (Section~\ref{arrow}), and conclude this perspective by comparing our argument with other fundamental principles proposed to explain biological complexification and the open-ended evolution (Section~\ref{discussion}).


\section{Open-ended biological complexity and evolutionary transitions}
\label{bio-view}

In this section we discuss various views on biological complexification. This process, observed in evolutionary dynamics over time, is  punctuated by major evolutionary transitions and ``coding thresholds''. In reviewing established approaches, we highlight key challenges and common features shared by these perspectives.

\subsection{Major evolutionary transitions}

The Second law of thermodynamics states the impossibility of fully transforming disordered heat energy into coherent work without the expense of some additional resource. Importantly, the Second law implies that physical systems naturally decay towards less structured arrangements. 
When seen against this background, the spectacular evolution of living systems on Earth is even more remarkable, requiring explanations that are compatible with these limiting physical principles.

When explaining the progress of biological evolution, two main perspectives have been adopted: approaches that emphasise that evolution happens slowly and continuously, and approaches that state that evolution is mainly driven by abrupt  transitions~\cite{eldredge1972punctuated,rhodes1983gradualism}. Within the second family of approaches, it has been argued that evolution displays major transitions in terms of biological complexity~\cite{szathmary1995major}. 
In general, these transitions produce important changes in the way organisms exchange information, cooperate and coordinate with each other, and how they survive and replicate. These collectives can become so  tightly arranged --- via functional specialisation, which enhances the efficacy of cooperation while inducing mutual dependency~\cite{west2015major} --- that they start acting as `higher-order' units that effectively drive the selection process~\cite{rajpal2023quantifying}. 
In effect, by tying each gene’s replication to the survival of higher-order structures, selection favours genes that promote survival of the higher-order structure, which in turn enables greater division of labour and specialisation. 

When trying to characterise what these major evolutionary transitions have in common, one can see that they involve profound changes in what an individual is and how it preserves its properties to new generations. This is usually supported by the emergence of novel inheritance systems, involving new modes of storing, transmitting, and processing information --- sometimes referred to as new ``codes of life''~\cite{barbieri2008mechanisms,kun2021major}. Examples of this include replicating molecules which form cells as compartmentalised ``populations'' of molecules; independent replicators (hypothesised to be RNA) which may have formed chromosomes to reduce information loss during replication; emergence of the genetic code and translation machinery (with DNA used as genes and proteins as enzymes); prokaryotes evolving into eukaryotes with a membrane-bound nucleus which stores their genetic information, and so on, towards the evolution of multicellularity and eusociality, as well as language and sociocultural evolution~\cite{szathmary1995major,west2015major,kun2021major}.

In this way, natural evolution is thought to give rise to more elaborate arrangements of higher-order organisms, from eukaryotic cells to multicellular organisms and even collectives such as swarms and hives, supported by novel information-processing modes. 
However, there are more ``codes of life'' than major evolutionary transitions~\cite{kun2021major}, and several crucial questions remain unanswered:
Are there computational principles that drive these evolutionary transitions? What are the computational trade-offs that are being negotiated during a transition?

\

\boxsection{

{\textbf{Box 1: Genetic information and replication}}

\subsubsection*{Gene transfer}
Transmission of genes from one generation to the next, i.e., from parent to offspring (as a result of sexual or asexual reproduction), is referred to as vertical gene transfer (VGT)~\cite{bright2010complex}. Horizontal gene transfer (HGT) moves partial genetic information laterally across distantly related organisms in the same generation~\cite{keeling2008horizontal}. 

\subsubsection*{Genotype and phenotype}
The genotype contains the specific genetic information an organism carries in its DNA, encoded in sequences of nucleotides. The phenotype comprises the set of observable traits (physical, biochemical and behavioural) of the organism. Genetic instructions are used in development and functioning of a living organism: they are involved in construction of other components and copying itself.

\subsubsection*{Extended phenotype}
The extended phenotype encompasses all the effects that a genotype can have on the environment beyond the immediate organism itself. This includes modifications or structures an organism creates that affect its survival and reproduction, influencing other organisms in the ecosystem~\cite{dawkins2016extended}. 
Extended phenotype includes (i) behavioural innovations, e.g., a bird’s ability to construct complex nests; (ii) interactions with environment and other organisms, indirectly influenced by traits encoded in the genome, e.g., pheromone production in insects affects their mating, foraging and social behaviours; (iii) physical modifications of the organism's environment, e.g., beaver dams, spider webs, or ant pheromone trails arise from behaviours guided by genetic predispositions and environmental factors.

\subsubsection*{Gene expression}
Transcription and translation are two fundamental processes of gene expression through which the information encoded in DNA is used to produce functional proteins that contribute to an organism's phenotype. Transcription is the process of making RNA copies of the genetic protein information encoded in DNA, and translation is the decoding of instructions for making proteins. 

}

\subsection{Coding thresholds and phase space expansions}
\label{threshold}

It has been argued that major evolutionary transitions  overcome specific ``coding thresholds'', with the saltations opening a novel evolutionary phase-space with qualitatively new possibilities to handle information~\cite{woese2004new,kun2021major}. For example, when discussing the emergence of DNA, Woese argued that
\begin{quote}
``Somewhere along the line there had to have occurred a saltation that we could call the “coding threshold,” where 
the capacity to represent nucleic acid sequence symbolically in terms of a (colinear) amino acid sequence developed, a development that would generate a truly enormous new, totally unique evolutionary phase space.''\cite{woese2004new} 
\end{quote}
Furthermore, \citet{woese2004new} suggested that another such saltation --- which he described as ``Darwinian threshold'' --- may have been crossed when cells went from an initial arrangement where their evolutionary dynamics were dominated by horizontal gene transfer (HGT) to a new arrangement dominated by vertical inheritance. This turned evolution from a communal process (in which evolving cells maintained no stable genealogical information) to one where stable organismal lineages could develop, preserve their genetic makeup, and coexist~\cite{woese2004new} (see subsection~\ref{HGT-innov}). 

Analogously, a ``language threshold'' may have been crossed when early human groups developed language (another ```codes of life''~\cite{barbieri2008mechanisms,kun2021major}), which enabled a new way to inherit knowledge and significantly speed-up adaptation. ~\citet{woese2004new} elaborated on this conjecture, noting that one common feature of higher-order arrangements --- such as multicellularity or human language --- is their enhanced communication ability, which enables a kind of `interaction at a distance'. This insight reinforces the idea that higher-order structures are able to synergistically access an extended problem-solving space, making such structures non reducible to a mere aggregation of information-processing subunits.

Unfortunately, however, the investigation of major evolutionary transitions is hindered by a chicken-or-egg causal dilemma, highlighted by various error threshold paradoxes (e.g., Eigen paradox~\cite{eigen1971selforganization}), as it is often difficult to explain how a more complex structure can evolve if the evolutionary benefits can be realised only at the higher level,  
This challenge was eloquently expressed in 1904 by a pioneer of genetics Hugo de Vries: ``natural selection may explain the survival of the fittest, but it cannot explain the arrival of the fittest'', and remains a subject of active research. For example, \citet{wagner2014arrival} explored how innovation and adaptability drive evolutionary success, arguing that the ability of organisms to generate new traits and capabilities is crucial for their long-term survival and thriving, and highlighting that constraints, rather than being purely limiting, can actually drive innovation.

An intriguing conjecture implicated genetic parasites, e.g., plasmids or viruses, as major catalysts for evolutionary transitions. Such parasites may (i) necessitate the formation of more complex structures helping the host to combat the parasitic threats and win in the parasite-host arms race, while (ii) providing building blocks --- via a parasite-enabled gene transfer --- for constructing the higher levels of organisation: 
\begin{quote}
``only increasing organizational complexity, e.g. emergence of multicellular aggregates, can prevent the collapse of the host–parasite system under the pressure of parasites''~\cite{koonin2016viruses}.
\end{quote}
In an abstract sense, this hypothesis suggests that the underlying inconsistencies and tensions may provide a dialectic impetus for evolutionary innovations, forcing the construction of new phase-spaces. Nevertheless, formulation of general computation-theoretic principles underpinning an open-ended phase-space expansion remains elusive.

\subsection{Innovation-sharing at the coding threshold}
\label{HGT-innov}

\citet{vetsigian_collective_2006} considered  evolution of the genetic code during early life, that is, during the stages preceding the emergence of vertical genealogical descent. They argued that early evolutionary dynamics of cell-like entities involved communal descent mediated by HGT. 

A key insight of their analysis is that the  genetic code emerging in this communal world was not only needed to encode amino acid sequences in the genome, but also served as an \textit{innovation-sharing protocol}. They conjectured that “early life did not require a refined level of tolerance”, and hence, the emerging innovation-sharing protocol allowed for imprecise copies created by an ambiguous translation without a unique mapping between codons and amino acids. Nevertheless, this dynamic produced a (nearly) universally common mechanism for encoding information, the {\it universal genetic code}:
\begin{quote}
``HGT of protein coding regions and HGT of translational components ensures the emergence of clusters of similar codes and compatible translational machineries. Different clusters compete for niches, and because of the benefits of the communal evolution, the only stable solution of the cluster dynamics is universality''~\cite{vetsigian_collective_2006}.
\end{quote}
The study concluded that once the universal genetic code emerged, the genetic complexity is likely to grow exponentially. This, in turn, leads --- via another (Darwinian) transition --- to dominance of the vertical (individual) descent, and hence the Darwinian evolution~\cite{vetsigian_collective_2006}.

\subsection{Division of labour increases fitness and the dimensionality of phenotype space} 

Generally, biological fitness consists of two components -- fecundity (reproduction) and viability (survival), and trade-offs between the two attributes shape diverse life-histories. In the context of clonal multicellularity, or a colony of genetically related cells, such trade-offs provide a selective pressure towards cell specialisation since group fitness can potentially be greater than the average of individual cells~\cite{michod2006group}. For example, a microtubule organising centre (MTOC) can typically be utilised to either support flagella for motility or cell division, but not both simultaneously. Hence, single-celled organisms typically separate reproduction and motility phases temporally. In multicellular Volvox this trade-off is thought to have led to the evolution of a spatial distinction between two groups of cells, corresponding to a germ/soma separation~\cite{michod2007evolution}, and a similar scenario may have occurred during animal evolution~\cite{king2004unicellular}.

\

\boxsection{

{\textbf{Box 2: Inclusive fitness}}

\subsubsection*{Darwinian fitness}
Darwinian fitness is a quantitative measure of reproductive success. More specifically it measures the average contribution to the gene pool of the next generation from an individual, classified by genotype. It depends on, for a given environment, the relative probability of survival and rate of reproduction.

\subsubsection*{Inclusive fitness}
Inclusive fitness is similar to Darwinian fitness but is helpful in explaining social interactions since it explicitly recognises that an individual can influence their genetic contribution to the next generation via their influence on individuals other than themselves. It takes into account the influence of such interactions on both the individual and the individual being influenced as well as the degree of shared genetic information between the two.

\subsubsection*{Multicellular organisms}
Multicellular organisms consist of more than one cell. Most multicellular organisms develop clonally, that is from the division of a single cell. This in turn means that most cells in such organisms are genetically identical (exceptions are caused by mutations during development or by genetic recombination in the case of germ cells). Less commonly, multicellular organisms form through aggregation, where single cells may come together to form the organism in response to environmental cues. In this case the cells involved are usually more genetically heterogeneous.
}

\vspace{1cm}

There are many other trade-offs beyond the general categories of survival vs reproduction that are relevant to a division of labour. For example, oxygen sensitive processes such as nitrogen fixation~\cite{fay1992oxygen} are separated spatially from oxygenic photosynthesis in filamentous cyanobacteria while the exchange of metabolites between cells benefits the whole~\cite{yoon2001pats}. Considering extant complex organisms and their many distinct cell types, the concept of division of labour can obviously be extended to a great variety of cellular functions. In general, a division of labour within multicellular organisms can be viewed as an increase in the dimensionality of phenotype space, since spatial separation enables otherwise incompatible processes to occur:
\begin{quote}
    ``Mathematically, the evolutionary advantage of the division of labour in aggregate forms can be viewed as the emergence of new, higher fitness maxima when the dimensionality of phenotype space is increased. The new fitness maxima are not a direct consequence of aggregation, but are based on the interaction between aggregated individuals that engage in the division of labour.''~\cite{ispolatov2012division}.
\end{quote}
Thus, given complementary cells states exist even in single-cell colonial communities, a division of labour may have provided an advantage to cell aggregates during the earliest stages of multicellular evolution~\cite{ispolatov2012division,tong2022selective}.

\subsection{Higher-level agents operate with extended patterns of information}

A higher-level aggregate organisation emerging out of interactions among the lower-level components may influence its constituent parts within a complex feedback loop~\cite{hoel2018agent}. 
\citet{mcmillen2024collective} investigated the competition  between collective and individual biological levels, in which ``the behavior of subunits percolates up toward adaptive processes at higher levels'', while ``higher levels of organization constrain and facilitate the behavior of their parts''~\cite{mcmillen2024collective}. In general, the notion of a biological individual, the question of group selection and the nature of competition between collective and individual levels should be interpreted from a temporal, ``diachronic'', perspective~\cite{okasha2006evolution}. This view allows us to consider intermediate stages in evolutionary transitions which create the \textit{potential} for ``conflict between levels of selection, for selection between the smaller units may disrupt the well-being of the collective''~\cite{okasha2006evolution}.

In adopting this approach, \citet{mcmillen2024collective} argued that one of the advantages of multiscale organisation is the capacity of collective agents to create a novel problem-space and modify the energy landscape available at the higher level, which allows the lower-level components to operate more efficiently~\cite{mcmillen2024collective}.
In this interpretation an energy function represents an optimisation problem --- an approach which traces back to the study of neural circuitry by~\citet{hopfield1985neural}. This seminal work demonstrated that interconnected networks of analog neurons can be computationally effective in solving hard optimisation problems, with the optimal solution corresponding to the lowest energy state of the network's energy function (defined in terms of neuronal voltages), i.e., its most stable state. In general, as has been pointed out by \citet{watson2011global},
\begin{quote}
``...in systems built out of the superposition of many low-order constraints, low-energy (high-utility) attractors necessarily have large basins of attraction... So, the better the attractor, the more it is visited, thus the more it is enlarged by learning, and the more it is visited in the future, and so on.”~\cite{watson2011global}.
\end{quote}
Adopting a broader definition of ``collective intelligence'' for hierarchical biological systems which consist of many interconnected parts (e.g., gene regulatory networks, cells, etc.) utilises the analogy with neural networks, suggesting that evolution involves altering those connections in a way similar to the training of a neural network (see also subsections~\ref{tension} and~\ref{self-model}). When the higher-order system attains a lower energy by modifying the energy landscape, it allows the collective to achieve a higher utility.

In information-processing terms, the access to a novel problem-space means that ``collective intelligence of competent parts'' is able to propagate information across scales, exhibiting an integrated problem-solving capacity, so that the higher-level agents are able to make decisions based on \textit{extended patterns of information}~\cite{mcmillen2024collective}. For example, gene expression of the frog embryo is influenced by the spatial voltage differences across its brain regions (i.e., by a group-level pattern), rather than by the absolute values of  individual cells~\cite{pai2015endogenous,pai2015local,mcmillen2024collective}. As a result, the dynamics is able to visit larger spatial areas, approaching the higher-utility (i.e., lower-energy) attractors with larger basins of attraction.

\subsection{Tensions between individual and group interests}
\label{tension}

It has been argued that the emergence of higher-level organisation generates a tension between levels. As noted by ~\citet{szathmary1995major}: “entities that were capable of independent replication before the transition can replicate only as part of a larger whole after the transition”.
This notion has been further developed by~\citet{watson2016can} (see also \cite{watson2016evolutionary}), who pointed out that ``in evo-ego, correlations change the evolutionary unit (such that multiple, previously separate units become a new single unit at a higher level of organisation)''. This analysis is closely related to the study of ``collective intelligence'' by~\citet{mcmillen2024collective}, and reinforces the observations that a higher-level organisation is qualitatively different and expands the phase-space of possibilities. 

Importantly,~\citet{watson2016can} explored the ``evo-ego'' relationship, highlighting a tension between individual and group interests: ``individual-level selection will oppose the creation and maintenance of adaptations that enforce selection at the group level''.  They emphasised that the transitions in individuality  essentially create new evolutionary units and may contribute to \textit{the evolution of evolvability}~\cite{dawkins2019evolution}, by evolving new mechanisms of inheritance or reproductive codispersal. Crucially, this approach identified a potential tension: 
\begin{quote}
``...if individual and group interests are aligned then selection applied at the group level does not alter evolutionary outcomes, and if individual and group interests are not aligned then individual-level selection will oppose the creation and maintenance of adaptations that enforce selection at the group level~\cite{okasha2006evolution}.''~\cite{watson2016can}. 
\end{quote}

\citet{watson2016can}, as well as \citet{watson2016evolutionary}, examined this tension, aiming to explain how evolution at one level of biological organisation (e.g., individual cells) can systematically generate reproductive structures non-trivially adapting at a higher level of organisation (e.g., multicellular organisms), even ``before that level of adaptation exists?''~\cite{watson2016can}. The proposed approach  --- evolutionary connectionism --- is discussed in sections~\ref{self-model}--\ref{connect-scaffold}.

\subsection{Ecological scaffolding}
\label{scaffold}

An alternative approach to explaining evolutionary transitions in individuality is offered by 
\citet{black2020ecological}, who pointed out that these transitions are immediately related to the emergence of biological complexity. In trying to analyse and clarify the conditions favouring emergence of collective-level reproduction (e.g., multicellular life), they proposed the concept of “ecological scaffolding”. Specifically, they modelled that, given an ecological structure of distributed and dispersing resources, a division of labour would allow individual cells to  ``participate directly in the process of evolution by natural selection as if they were members of multicellular collectives''~\cite{black2020ecological}.

\section{Beyond computational undecidability: ``breaking'' the limits of computation}
\label{comp-view}

In this section, we turn our attention to computation-theoretic approaches that can provide formal means to analyse an open-ended process of complexification. This computation-theoretic background will be used in subsequent sections to describe novelty generation during open-ended evolution (discussed in Section \ref{bio-view}).

In general, the problems or functions that cannot be solved or computed by any algorithm are referred to as \textit{incomputable}.
For example, the Halting Problem, a classic problem in the theory of computation, asks whether it is possible to create an algorithm that can determine, for any arbitrary program and input, whether that program will eventually halt (terminate) or continue running indefinitely~\cite{turing_halting_1937}. The halting problem is undecidable: no general algorithm exists that solves the halting problem for all possible program–input pairs. 
Likewise, Gödel showed that in any sufficiently powerful formal system, there are true statements that cannot be proven within the system \cite{godel_uber_1931}. 

\subsection{Self-reference and diagonalisation arguments}
\label{diag-prelim}

A self-referential expression is one that refers to itself literally, e.g., the phrase ``this sentence'' in the liar sentence ``this sentence is false'' or via its referents i.e., distinct expressions that \textit{denote}, \textit{name}, or \textit{encode} the original expression~\cite{sep-self-reference}. A notable example of self-reference by referents is Gödel numbering, where statements about natural number arithmetic are uniquely assigned a natural number themselves --- so self-reference arises when statements about natural number arithmetic are applied to their own Gödel number \cite{kripke-self-reference}.

\

\begin{figure}[!h]
    \centering
\includegraphics[width=1.0\columnwidth]{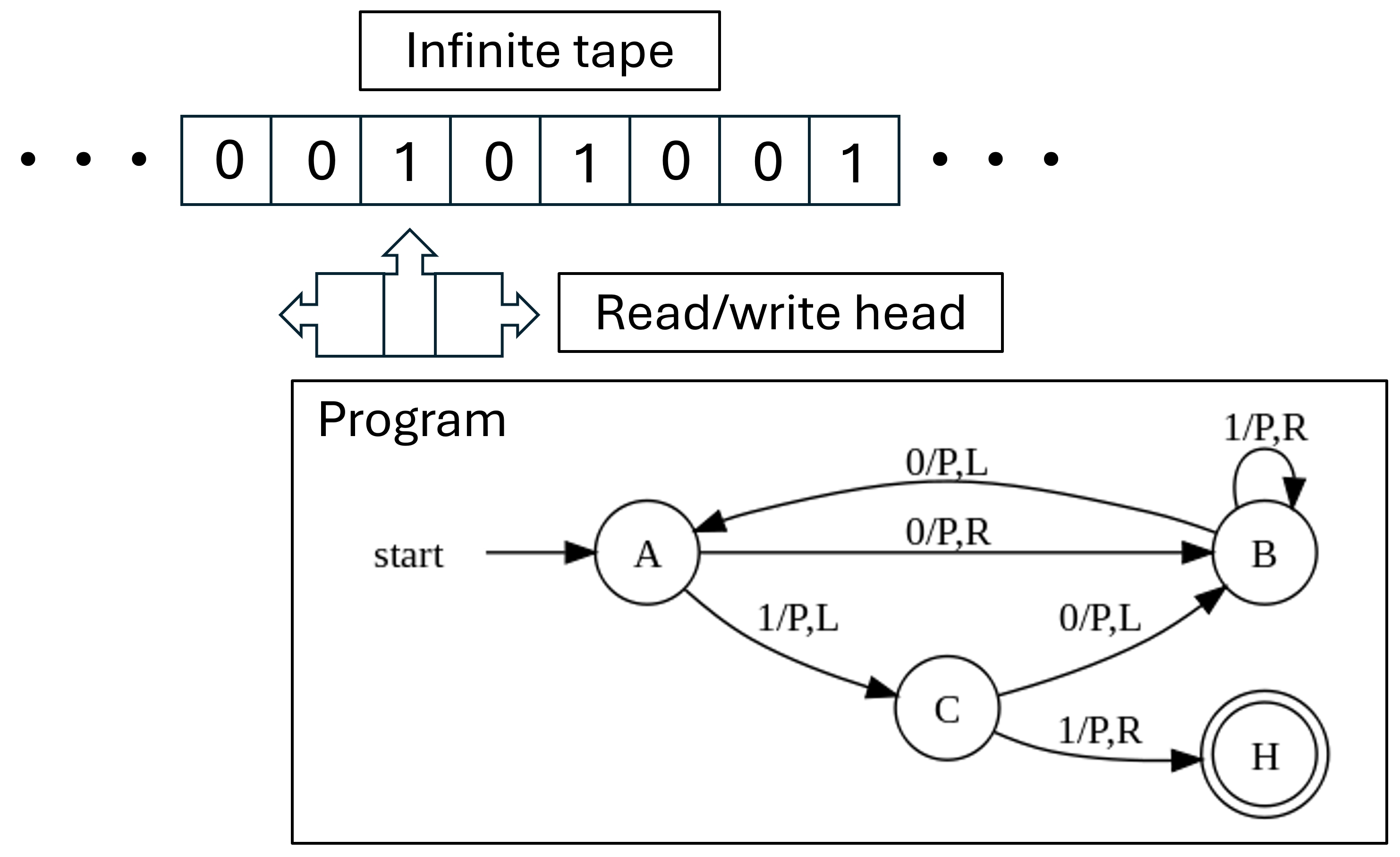}
    \caption{An example finite-state specification of a Turing Machine. States include A, B, C and halting state H. An arrow label (e.g., 0/P,R) specifies the tape symbol (e.g., symbol 0) that upon reading triggers a particular transition to another state, followed by the action, e.g., print (P) and move tape to the right (R)~\cite{wiki-TM}. }
    \label{fig:turing-machine}
\end{figure}

\begin{figure}[!h]
    \centering
\includegraphics[width=1.0\columnwidth]{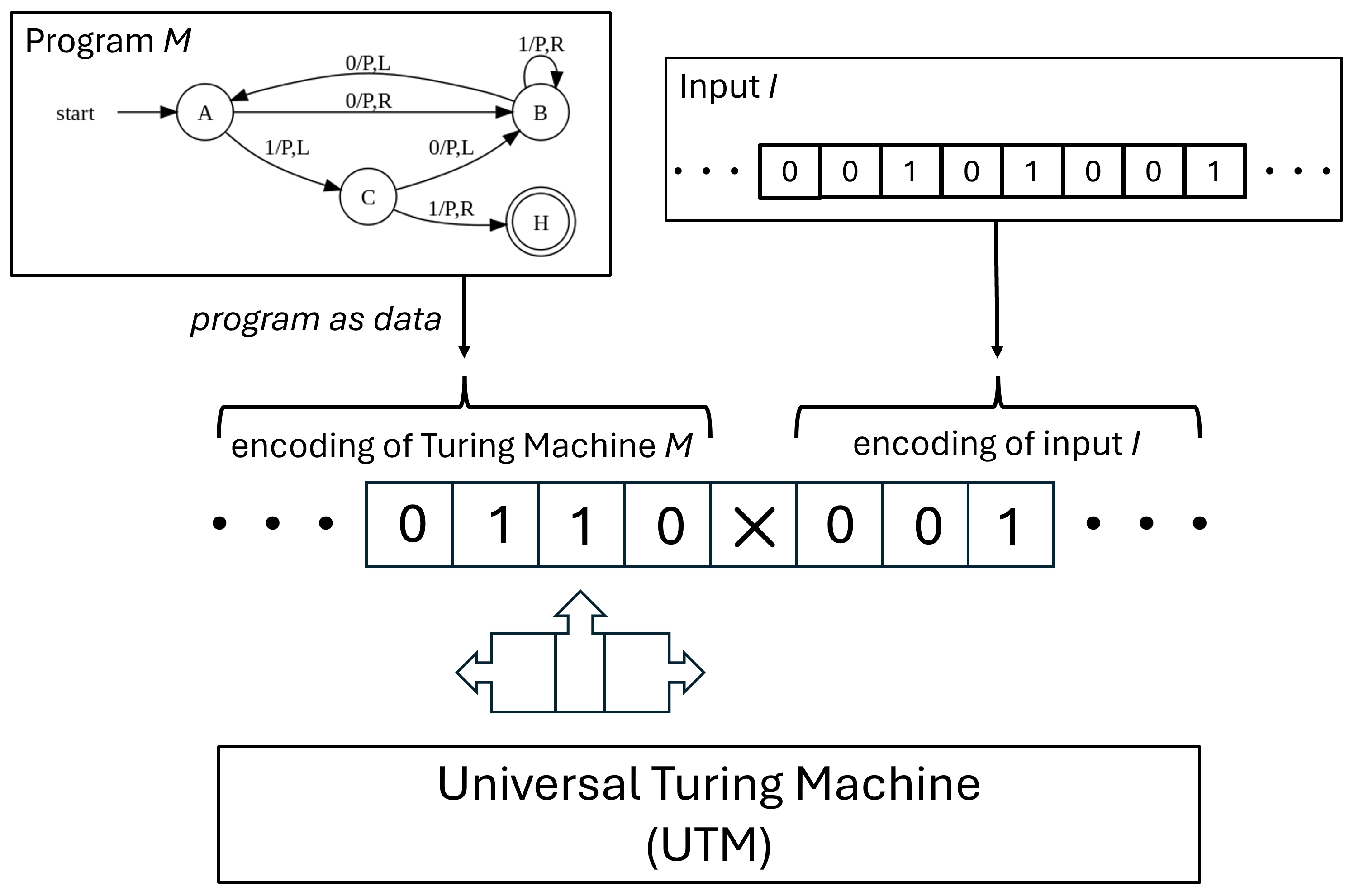}
    \caption{Universal Turing Machine (UTM) simulating Turing Machine $M$ on input $I$. An encoding of program $P$ converts it into input data which is separated on the tape from the input $I$ by suitably chosen separator symbol(s) $\times$.}
    \label{fig:utm}
\end{figure}

\boxsection{

{\textbf{Box 3: What is computation?}}

\subsubsection*{Algorithms}

Computation is defined as a process of executing a series of \textit{well-defined} instructions such that given an input state, its execution may satisfy a termination condition, which produces a corresponding output state.\\ 

An algorithm is a finite sequence of mathematically rigorous instructions, typically used to perform a computation.

\subsubsection*{Turing machines}

A model of computation, e.g., Turing machine, is an
abstract model which describes how an output of a mathematical
function is computed given an input.
Specifically, Turing machines model a mechanism that operates on an infinitely long input tape of discrete symbols, using a head that can read or write a symbol in a given position, store a state and move in either direction of the tape. Turing Machines include transition rules with respect to the state and symbol at its head, and terminating transitions known as halting conditions (see Fig. \ref{fig:turing-machine}). The final state of the tape corresponds to the computation's output.\\

Turing machines have become synonymous with general-purpose computers  because \textit{Universal Turing Machines} (UTM) can be constructed such that the specifications of any Turing Machine (any algorithm) and input can be encoded as a UTM input, whereby the UTM's execution produces the output that matching the original Turing Machine applied to the original input (see Fig. \ref{fig:utm}).

\subsubsection*{Information processing}

It is useful to distinguish computation from ``information processing'', the latter referring to the manipulation or transformation of data or information by a system, such as encoding, compression, storage,  transmission, decoding, search, pattern recognition, retrieval, etc. While information processing typically transforms input data into an output, in general, it does not have to produce a meaningful output and may continue without a termination condition. Hence, information processing differs from computation, which is a process with well-defined algorithm and termination (output) states. (See \cite[ch. 12]{mitchell2009complexity} for further discussion).
}

\

\

Self-referential statements have historically been employed to prove results related to incomputability, using what are known as \textit{diagonalisation} arguments. A canonical setting for diagonalisation arguments are expressions $E_i$, referents $\NGodel{E_j}$ and properties $P_{ij}$ arising out of the application of expressions $E_i$ to referents $\NGodel{E_j}$, which we can visualise as the following grid: 
\begin{equation}
    \begin{tabular}{c | *{6}{c}}
        \multicolumn{1}{c}{} & \multicolumn{1}{c}{$\NGodel{E_0}$} & \multicolumn{1}{c}{$\NGodel{E_1}$} & \multicolumn{1}{c}{$\NGodel{E_2}$} & \multicolumn{1}{c}{$\cdots$} & \multicolumn{1}{c}{$\NGodel{E_k}$} & \multicolumn{1}{c}{$\cdots$} \\
        \cline{2-7}
        $E_0$ & $P_{00}$ & $P_{01}$ & $P_{02}$ & $\cdots$ & $P_{0k}$ & $\cdots$ \\
        \cline{2-7}
        $E_1$ & $P_{10}$ & $P_{11}$ & $P_{12}$ & $\cdots$ & $P_{1k}$ & $\cdots$ \\
        \cline{2-7}
        $E_2$ & $P_{20}$ & $P_{21}$ & $P_{22}$ & $\cdots$ & $P_{2k}$ & $\cdots$ \\
        \cline{2-7}
        $\vdots$ & $\vdots$ & $\vdots$ & $\vdots$ & $\ddots$ & $\vdots$ & $\vdots$ \\
        \cline{2-7}
        $E_k$ & $P_{k0}$ & $P_{k1}$ & $P_{k2}$ & $\cdots$ & $P_{kk}$ & $\cdots$ \\
        \cline{2-7}
        $\vdots$ & $\vdots$ & $\vdots$ & $\vdots$ & $\vdots$ & $\vdots$ & $\ddots$ \\
    \end{tabular}
    \label{diagonalization}
\end{equation}
Incomputability is then represented by finding combinations of expressions $E_i$ and referents $\NGodel{E_j}$ whose property $P_{i,j}$ can not be determined without a contradiction. These combinations typically arise out of constructions on properties of the form $P_{kk}$, which are considered to be self-referential as they represent the application of an expression to its own referent.

To illustrate how diagonalisation arguments work, let us consider Cantor's Theorem. This result states that the subsets of the natural numbers (e.g., $\{2, 5, 1\}$) are uncountable, i.e., that any list of subsets $E_0, E_1, E_2, \ldots $ (indexed by the natural numbers) is incomplete~\cite{cantor}. The argument begins by assuming that an arbitrary such list is complete. For such a list, one builds an expression-referent grid as follows: expressions $E_i$ correspond to the subsets, referents $\NGodel{E_i}$ indicate the corresponding index (i.e. $\NGodel{E_i} \coloneqq i$), and properties $P_{ij}$ are defined as $P_{ij} = 1$ if $i$ is an element of $E_j$ and $0$ otherwise. The second step of the argument to prove incompleteness is to find $i$ and $j$ such that $P_{ij}$ can not be determined. To do this, one uses the diagonal to construct the set $\{k\in\mathbb{N}\ |\ P_{kk} = 0\}$, i.e., each number $k$ when considered as a referent is not a member of the expression referenced by it. Because this is a subset of the natural numbers, it must be an expression within our list, say $E_x$. However, any attempt to evaluate $P_{xx}$ leads to a contradiction: assuming the referent $x$ is not in $E_x$ leads to the conclusion that $x$ \textit{does} belong to $E_x$, and \textit{vice versa}. Therefore, we use the incomputability of $P_{xx}$ to conclude that our initial list could not have been complete.

Another diagonalisation argument can be used to show that the real numbers are uncountable. Using an expression-referent grid formed by enumerable rows of infinite sequences of binary digits, Cantor demonstrated that while there are countably many columns (corresponding to the natural numbers), there are in fact uncountably many rows (representing the real numbers) --- i.e., that such grid is not a square. In doing so, Cantor used a diagonalisation argument constructing a real number $E_v$ as a sequence of binary digits $P_{vj}$ such that each digit is complementary to its diagonal counterpart $P_{jj}$ (essentially, swapping zeros and ones in $P_{jj}$). By definition, the constructed sequence $E_v$ is different from any countable sequence $E_k$, disagreeing with it on at least one digit. 

\subsection{Discrepancy between expressions and referents}
\label{expr-ref}

Importantly, diagonalisation arguments reveal and exploit a fundamental discrepancy between the space of expressions and the space of referents: in general, there are \textit{more} expressions than referents. In other words, it is not possible to construct a one-to-one correspondence between these two spaces. 
For instance, by showing that the real numbers are \textit{uncountable}, Cantor proved that there are \textit{more} real numbers than natural numbers --- despite both sets being infinite. 
The fundamental expression-referent discrepancy forms a key part of our analysis. It is also described in \ref{diag} in the context of the halting problem.

\subsection{Incomputability and oracle machines}
\label{open-ended-incomp}

\begin{figure}[!h]
    \centering
    \includegraphics[width=1.0\columnwidth]{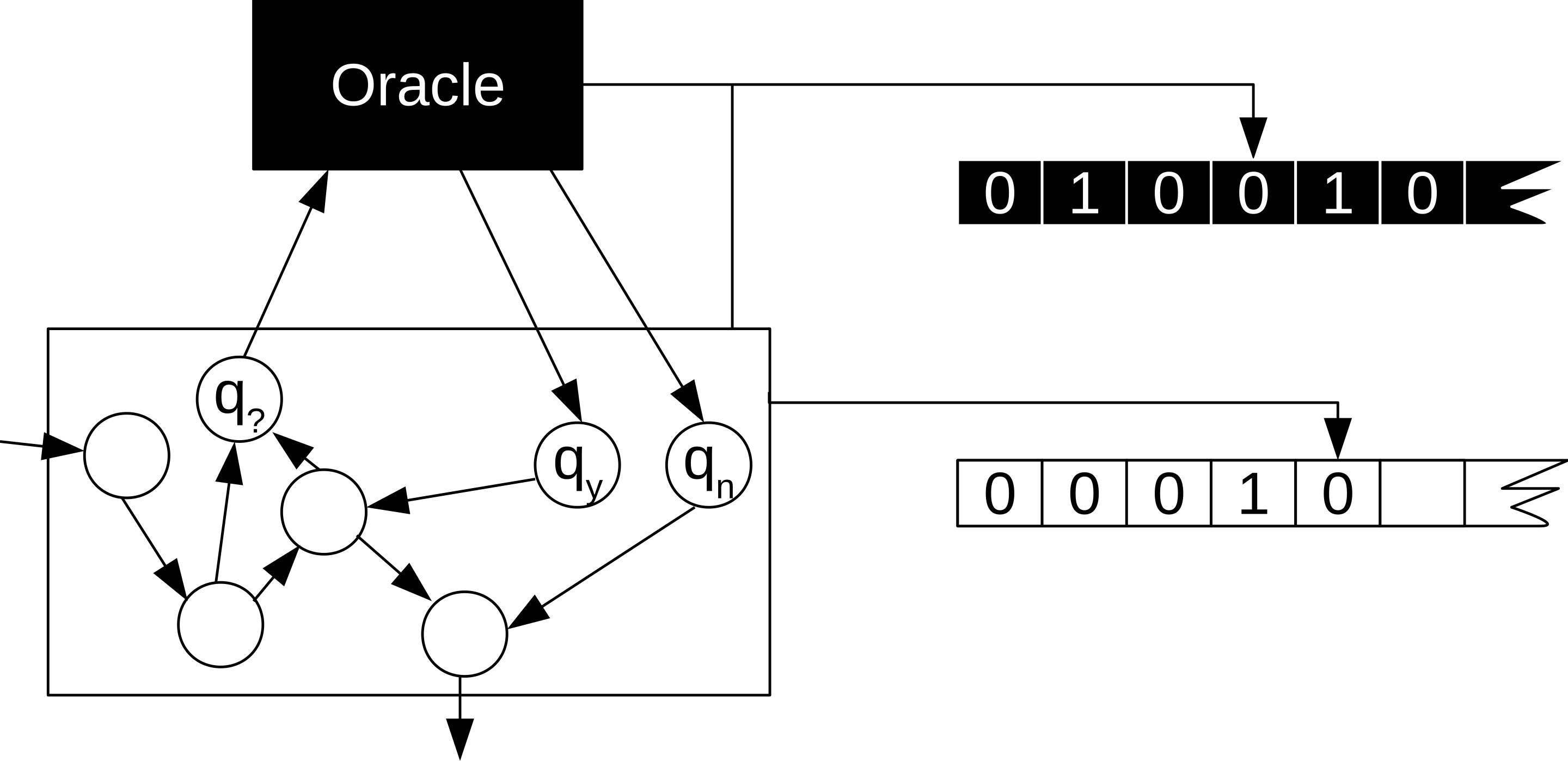}
    \caption{Turing oracle. Wikimedia Commons, licensed under the Creative Commons Attribution-Share Alike 4.0 International license.}
    \label{fig:turing-oracle}
\end{figure}

As somewhat ironically noted by~\citet{soare2009turing}, ``the field of computability theory deals mostly with \textit{incomputable}, not computable, objects''. 
Crucially, the incomputability of a given problem is considered relative to a given computational system, and in certain cases it can be overcome by considering another system that extends the original system in some way.   
For example, decision problems undecidable by Turing machines can be decided by a Turing machine extended with an additional component known as the \emph{oracle}~\cite{turing_systems_1939}. An oracle is a black-box capable of providing, in a single operation, an answer for any instance of the corresponding decision problem, as illustrated by Fig.~\ref{fig:turing-oracle}. Adding an oracle to a Turing machine is analogous to adding a new, independent, axiom to a formal logical system, so that the extended system is able to prove more than it could previously do~\cite{copeland2008mathematical}.

As a result of such extension, an \textit{oracle machine} --- a Turing machine connected to an oracle --- can simulate any Turing machine, producing the same output, given the same input, as the simulated Turing machine~\cite{soare2009turing}.
\citet{hofstadter_go_1980} informally referred to this way of resolving paradoxes as ``Jumping Out Of The System” (JOOTSing). However, such a resolution generates new undecidable problems within the extended (Turing machine plus oracle) system, leading to an open-ended process described in the next subsection. 

\subsection{Open-ended computational meta-simulation}
\label{ordinal}

In general, the idea of resolving undecidability by extending bounds of the computational system is developed within the field of recursion theory. In particular, the G\"odel--Turing--Post recursion-theoretic framework proposed various methods of constructing extensible ordinal or recursively generated logics \cite{godel_uber_1931,turing_computability_1937,turing_halting_1937, turing_systems_1939,post_recursively_1944, soare2009turing}.

\begin{figure}[!h]
    \centering
    \includegraphics[width=1.0\columnwidth]{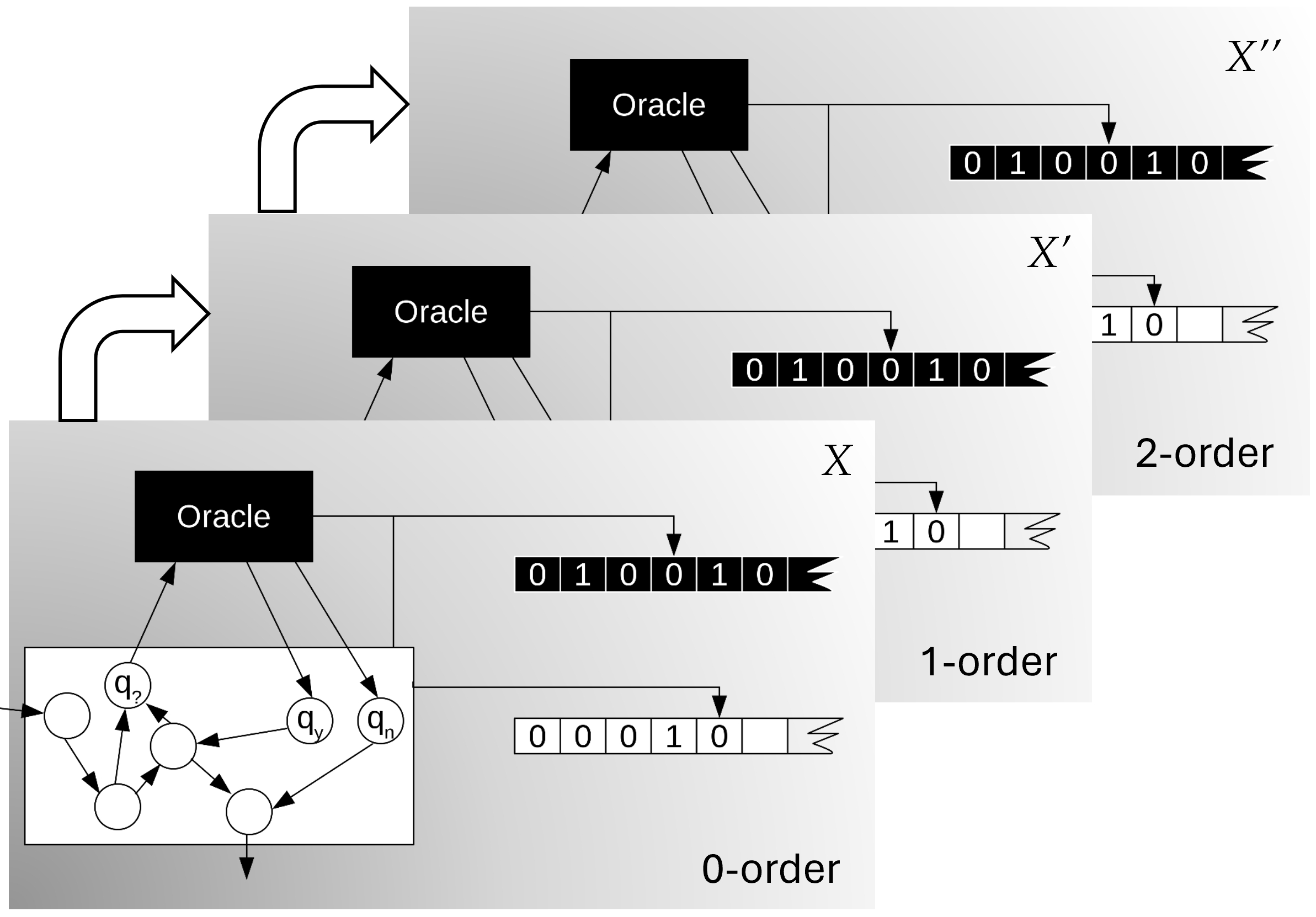}
    \caption{Open-ended sequence of $\alpha$-order oracle machines, with the corresponding Turing jump operations assigning successively harder decision problems $X, X', X''$,..., in an open-ended way. }
    \label{fig:alpha-oracles}
\end{figure}

\citet{turing_systems_1939} proposed a way of overcoming the halting problem  by providing a means to source an answer beyond the system boundary, that is, by considering an $\alpha$-order oracle machine specified for a particular level. In general, one may consider a sequence of $\alpha$-order oracles machines with increasing orders $\alpha$, each of which (for $\alpha > 0$) is capable of resolving undecidable halting problems of lower levels~\cite{turing_systems_1939,penrose1994shadows}, as illustrated in Fig.~\ref{fig:alpha-oracles}.

The $\alpha$-order oracles machines can be related by an operation known as the ``Turing jump''.
Formally, the ``Turing jump'' is an operation that assigns to each decision problem $X$ a successively harder decision problem $X'$ such that $X'$ is not decidable by an $\alpha$-order oracle machine with an oracle for $X$. Crucially, the Turing jump of $X$ generates an $(\alpha+1)$-order oracle to the halting problem for $\alpha$-order oracle machines with an oracle for $X$.

The boundaries of lower-order systems are expanded by adding specific ``novelties'' generated by an interaction with the corresponding oracle. In describing this continual iterative process, \citet{turing_systems_1939} drew an analogy with ordinal logics of \citet{godel_uber_1931}. Ordinal logic is associated with an ordinal number by recursively adding elements to a sequence of previous logics:
\begin{quote}
``The well-known theorem of Gödel (1931) shows that every system of logic is in a certain sense incomplete, but at the same time it indicates means whereby from a system $L$ of logic a more complete system $L’$  may be obtained... A logic $L_{\omega}$ may then be constructed in which the provable theorems are the totality of theorems provable with the help of the logics $L$, $L_1$, $L_2$,...''~\cite{turing_systems_1939}.
\end{quote}

This view was further developed by \citet{post_recursively_1944} using the recursively enumerable sets. In demonstrating that every  recursively generated logic may be extended, \citet{post_recursively_1944} proposed the concept of relative computability: degree of unsolvability, or the Turing degree, of a set of natural numbers measures the level of algorithmic unsolvability of the set. Consequently, \citet{post_recursively_1944} proposed a hierarchy of degrees of unsolvability, where each degree represents the extent to which a set is unsolvable or incomputable. Each level of the hierarchy corresponds to a different degree of computational complexity, with sets at higher levels being more incomputable than those at lower levels.

Importantly, each extended system $X^{(\alpha)}$ comprising an $\alpha$-order oracle machine ($\alpha > 0$) is able to simulate a lower-level $X^{(\alpha-1)}$ system. Arguably, one may consider a higher-order $\alpha$-order oracle machine as a meta-simulator, i.e., a simulator of nested, lower-order, simulations (down to the $0$-order oracle machine that can simulate any Turing machine). Thus, we suggest that when a higher-order system  simulates the lower-order computation, providing answers to lower-level undecidable problems, it performs \textit{meta-simulation} of the lower-order systems. We shall refer to the successive construction of extensible computational systems and ordinal or recursively generated logics (within the G\"odel--Turing--Post recursion-theoretic framework) as \textit{open-ended meta-simulation}.

\section{Undecidability and open-endedness in dynamical systems}
\label{dynamical}

The third primary domain, considered as part of our background, is physical dynamical systems. The concept of undecidable dynamics has been defined in physical systems by some analogy with computational undecidability, and so we can expect to see emerging parallels between unpredictable dynamics, expanding phase-spaces and extensible computational
systems.

\subsection{Undecidability in physical dynamical systems}
\label{moore}

Physical dynamical systems are inherently nonlinear and time--continuous.
According to \citet{Moore1990}, physical dynamical systems with at least three degrees of freedom can be Turing equivalent. They exhibit a type of unpredictability that is qualitatively stronger than low-dimensional chaos, with undecidable long-term dynamics, even if the initial conditions are known exactly.
\citet{Bennett1990} qualifies this as follows:
\begin{quote}
    “For a dynamical system to be chaotic means that it exponentially amplifies ignorance of its initial condition; for it to be undecidable means that essential aspects of its long-term behaviour --- such as whether a trajectory ever enters a certain region --- though determined, are unpredictable even from total knowledge of the initial condition.”
\end{quote}
Thus, undecidability is a more extreme kind of unpredictability than chaos \cite{Bennett1995}.
One implication of this is that if the halting problem is viewed through the lens of dynamical systems, then its undecidability means that the long–term behaviour of the dynamics corresponding to a universal Turing machine is unpredictable, even if the initial conditions are known exactly.

\begin{figure}
    \centering
    \includegraphics[width=0.9\columnwidth]{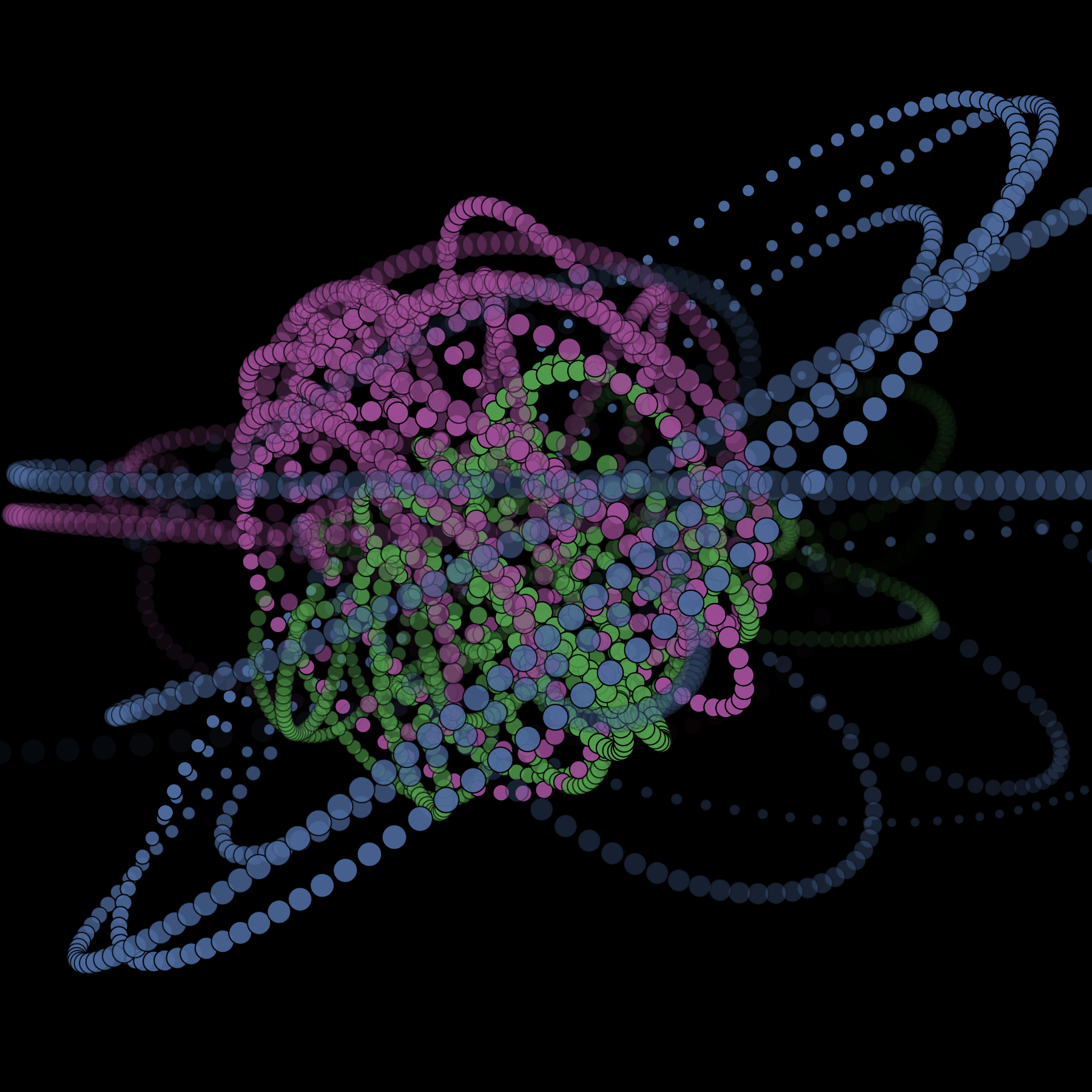}
    \caption{$n$-body simulation demonstrating unpredictability and undecidability of Newtonian dynamical systems, in this case comprised of $n=3$ stars of different masses (indicated by coloured trajectories) orbiting each other under the influence of gravity.}
    \label{fig:nbody}
\end{figure}

A relatively simple example offered by Moore is the motion of a particle in a 3D potential. He suggests, however, that the three-body problem ($n$-body more generally) presents more realistically complex dynamics for testing these ideas, in particular by demonstrating lack of scaling behaviour, irregular spectra of periodic points and attractor states, which are beyond the features that typically characterise chaotic dynamics in non-physical dynamical systems (e.g. the Lorenz 63 system, a mathematical model that only loosely approximates the complex dynamics of real weather systems).
As shown in Fig.~\ref{fig:nbody} for the classic 3-body problem in a Newtownian gravitational system, it is impossible to predict if or when one of the orbiting objects becomes gravitationally unbound (i.e. the halting condition), even if the initial conditions (position and velocity of each object) are known.

Physical dynamical systems (e.g. fluid flows, propagating waves), as described by space--time partial differential equations, may have universal computing abilities if they have sufficient dynamical variables to maintain logically distinct trajectories in the presence of external noise. 
\citet{Bennett1995} argued that computationally universal dynamical systems can be used to define physical complexity, the quantity that increases when a self-organising system organises itself, producing novel features beyond those resulting from long-term evolution. 
\citet{wolpert2000computational} derived specific impossibility results associated with physical computation. This work developed analogues of Chomsky hierarchy results about universal Turing Machines and the Halting theorem, showing, in particular, the impossibility of certain kinds of error-correcting codes.
These observations not only open a way to connect computation-theoretic and dynamical-systems views on complexification processes, but also suggests that novelty generation during an open-ended biological evolution may also be seen through the lens of expanding computational spaces. These cross-disciplinary perspectives are explored in more detail in Section~\ref{cross-disc}.

\subsection{Expanding phase-space}

In contrast to non-physical dynamical systems, physical dynamical systems are constrained by the laws of physics, which could be argued as restricting phase space expansion.
However, physical systems may also exhibit stochastic (non-deterministic) dynamics, which, while also obeying laws of physics (e.g. thermodynamics), introduces extra degrees of freedom (i.e. higher-order complexity) in addition to any deterministic dynamics, provided the stochasticity is not purely random, without any underlying structure or information content.

In general, dynamical systems (both physical and non-physical) may encounter different dynamical phase-space regimes, e.g. ordered, chaotic and edge-of-chaos, with different computational representations.
Chaotic dynamics, associated with an increase in phase space exploration, continuously generates information and the amount of information needed for prediction grows with time. As discussed above, chaotic dynamics are not necessarily computationally undecidable.
Edge-of-chaos dynamics, on the other hand, has been proposed as optimal for information processing and novelty generation \cite{langton1990computation}. In physical systems, edge-of-chaos is associated with the spontaneous emergence of long-range spatio-temporal correlations, effectively expanding dynamical phase-space, with recent studies (e.g. \cite{boedecker2012information,Hochstetter2021}) demonstrating enhancement in information processing, but only in terms of general properties or for relatively complex computational tasks rather than necessarily for any specific task. 
\citet{prokopenko_self-referential_2019} examined the link between edge-of-chaos and undecidable dynamics in the context of cellular automata. This is briefly introduced in section~\ref{GOL} and discussed in more detail in Section~\ref{undecidable-dynamics}.

\subsection{Game of Life}
\label{GOL}

\begin{figure}[!h]
    \centering
    \includegraphics[width=1.0\columnwidth]{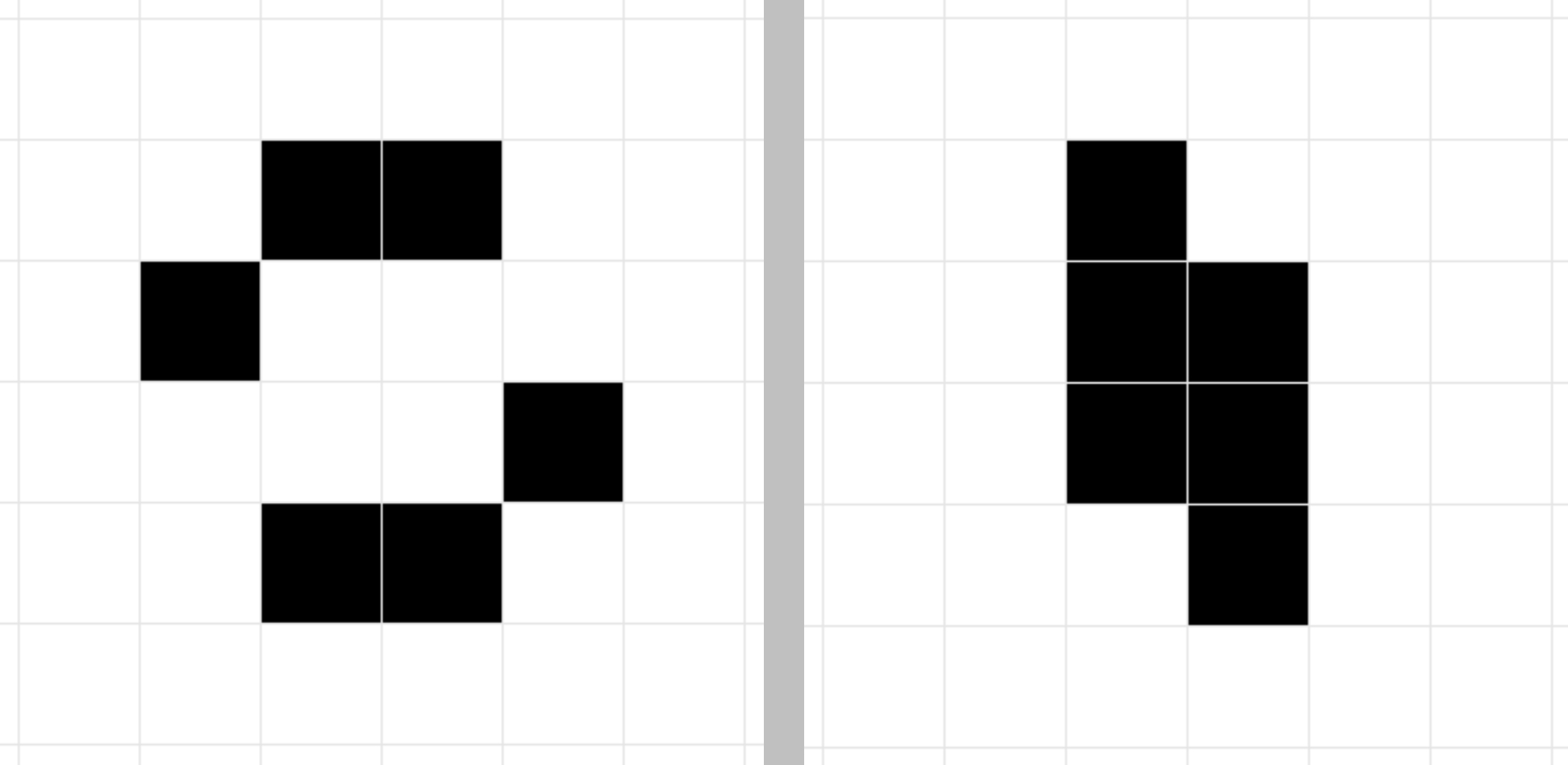}
    \caption{Two configurations of the oscillating polyomino pattern known in Conway's Game of Life as ``toad''. }
    \label{fig:toad}
\end{figure}

To illustrate how undecidable dynamics can be generated in a concrete system, let us introduce \textit{Conway's Game of Life} --- 
a well-known example of a discrete dynamical system~\cite{Gardner1970}. It is a two-dimensional cellular automaton (CA) (see~\ref{app-CA}), specified with a binary alphabet and a local update rule  defined for the Moore neighbourhood with 9 cells, incorporating:
\begin{enumerate}
\item  Deaths. Any live cell with fewer than two or more  than three live neighbours dies.
\item  Survivals. Any live cell with two or three live neighbours lives on to the next generation.
\item  Births. Any dead cell with exactly three live neighbours becomes a live cell.
\end{enumerate}
Game of Life dynamics produce coherent spatial patterns, including oscillators that repeat themselves after a fixed number of generations (e.g., see Fig.~\ref{fig:toad}), and gliders that move across the grid replicating their structure. It has been demonstrated that CA carry out distributed computation, with oscillators representing information storage (i.e., memory) \cite{lizier2012informationstorage}, gliders capturing information transfer (i.e., communications) \cite{lizier2008informationtransfer}, and glider collisions corresponding to information modification (i.e., processing) \cite{lizier2010informationmodification}, with the computational processes forming coherent information structures~\cite{lizier2012coherent}.

\

\boxsection{

{\textbf{Box 4: Arrows of time}}

There are several conceptually distinct arrows of time which are often interrelated in forming an understanding of the ``asymmetry'' of time, explaining why processes in nature appear irreversible and why we perceive time as moving forward from the past to the future. 

\subsubsection*{Thermodynamic arrow of time}

Thermodynamic arrow of time is defined by the second law of thermodynamics, which states that entropy in a closed system always increases over time, leading to the irreversibility of natural processes.  

\subsubsection*{Thermodynamic arrow of time and evolution}

\citet{blum2015time} discussed ``the relationship between time's arrow (the second law of thermodynamics) and organic evolution, exploring irreversibility and direction in evolution, and arguing that evolutionary patterns may be predetermined by thermodynamic processes.  Another hypothesis, Dollo’s law, suggests that once a complex trait or structure has been lost by an organism in the evolutionary process, it is unlikely to be regained in exactly the same form~\cite{collin2008reversing}. In other words, Dollo’s law  posits that evolution is not generally reversible, although reversible evolution has been observed on relatively short evolutionary timescales~\cite{clarke1985evolution}.

\subsubsection*{Cosmological arrow of time}

Cosmological arrow of time refers to the direction of time in which the universe is expanding, and is based on the observation that the universe is growing larger over time, as opposed to contracting.

\subsubsection*{Radiative arrow of time}

Radiative arrow of time aligns the direction of time with the way radiative processes, such as retarded electromagnetic radiation, the emission of light and sound waves, expand outward from their source,  from a higher energy state to a lower one.

\subsubsection*{Causal arrow of time}

Causal arrow of time reflects the principle that cause precedes effect: perceived events are ordered in a way that causes come before their effects. 

\subsubsection*{Quantum arrow of time}

Quantum arrow of time as defined in quantum mechanics relates time's direction to the collapse of the wave function. While the direction of time in a quantum system may be blurred due to uncertainty, when the system is measured, it transitions from a state of superposition to a definite state, which defines a temporal direction.
An insightful perspective which may characterise other time arrows was offered by Seth Lloyd: the arrow of time is an arrow of increasing correlations~\cite{lloyd1988black}.
}

\vspace{0.5cm}

It has been shown that computational power of Game of Life is equivalent to that of a universal Turing machine~\cite{Berlekamp1982}, and thus, the Game of Life dynamics may be undecidable under some conditions~\cite{prokopenko_self-referential_2019}. In other words, it cannot be determined, for all initial configurations of the Game of Life, whether or not they will reach some predefined final (termination) configurations (see \ref{app-CA}).

\vspace{-0.3cm}

\section{Cross-disciplinary perspectives on novelty generation}
\label{cross-disc}

In this section, we draw insights from several background studies across the three examined domains --- biological, computational and physical --- and explore possible connections among (a) open-ended biological complexity, including major evolutionary transitions (Section \ref{bio-view}); (b) open-ended computational meta-simulation, including successive resolution of lower-level undecidable problems (Section \ref{comp-view}); and (c) complex dynamical systems, including chaotic and strange attractors (Section~\ref{dynamical}). In doing so, we will try to examine how the ``evolving evolvability'' is related to emergence of  hierarchical structures comprising individual and collective information-processing elements, e.g., (a)  formation of new units of selection, comprising ensembles of pre-existing organisms, (b) computational novelty generation, and (c) emergence of self-replicating dynamic behaviours at the ``edge of chaos''.

\subsection{Chaos, incomputability, and undecidable dynamics}
\label{undecidable-dynamics}

Casti posited a link between the existence of chaotic dynamical systems and Gödel's incompleteness theorem \cite[p.317]{casti91} \cite[p.148]{casti94}. Casti argued that
\begin{quote}
``...the theorems of a formal system, the output of a UTM [Universal Turing Machine], and the attractor set of a dynamical process (e.g., a 1-dimensional cellular automaton) are completely equivalent; given one, it can be faithfully translated into either of the others'' \cite[p.317]{casti91}.
\end{quote}
By this correspondence Casti investigated dynamical systems that compute Kolmogorov complexity of numbers / programs, which is the length of the shortest programs that can compute them. Casti then applied Chaitin's (Incompleteness) Theorem \cite{chaitin74} on Kolmogorov complexity to prove such universal computing dynamical systems have strange attractors unreachable from any initial state, which Casti argued is equivalent to Gödel's (first) incompleteness theorem. Analogously, \citet{adams2017formal} pointed out that simple one-dimensional cellular automata possess many states that cannot be reached from any initial state by repeated application of any of the complete set of 256 evolution rules. Such states may, however, become accessible if the evolution rules themselves evolve with time.

Several fundamental relations between (i) formal systems, (ii) algorithms, and (iii) dynamical systems were identified by~\citet{prokopenko_self-referential_2019}. The comparative analysis identified three common factors implicated in the generation of undecidable dynamics within the three examined computational frameworks:
(i) the program-data duality (e.g.,  encoding of Turing machines as input data); (ii) access to an infinite computational medium (e.g., an infinite tape used by a Turing machine); and (iii) the ability to implement negation (e.g., the ability to flip \textit{accept} and \textit{reject} states).  

Considering ``undecidable dynamics'' in a broad context, \citet{prokopenko_self-referential_2019} generalised the concept of incomputability across several computational frameworks. For example, in formal systems, a \textit{dynamic process} can be identified with a \textit{proof}: a sequence of well-formed formulas derived by inference rules and starting from an axiom. Correspondingly, Gödel Incompleteness Theorem applicable to formal systems establishes the limits of \textit{provability} in axiomatic theories, capitalising on the observation that some formal sentences can neither be proved nor disproved (e.g., the famous Gödel sentence which encapsulated the Liar paradox). 

Similarly, for a Turing machine, a \textit{dynamic computational process} is provided by a \textit{sequence of machine states and tape patterns}, starting from some input. The halting Problem is the canonical example of computational \textit{undecidability}, capturing the fact that it cannot be established, for all program-input pairs, whether the computation reaches a predefined halting state or runs forever. 

Finally, one can consider an evolution of configurations --- a \textit{dynamic trajectory} --- within a dynamical system such as a Cellular Automaton, starting from an initial state. Given suitably defined termination conditions (e.g., testing for fixed points or limit cycles), one may attempt to establish whether the trajectory is reaching the predefined attractors or continues to unfold along the ``edge of chaos'', i.e., class IV Cellular Automata~\cite{wolfram1984universality}. The latter scenario can be identified with \textit{undecidable dynamics}; see also Section \ref{moore}.  

In summary, formal systems (e.g., logical systems), models of computation (e.g., Turing machines), and dynamical systems (e.g., Cellular Automata) are deeply related: these frameworks can produce universal computation and generate undecidable dynamics. This undecidability is manifested through three factors: self-reference, infinite computation, and negation~\cite{prokopenko_self-referential_2019}.

\subsection{Computational novelty generation}
\label{comp-novelty-gen}

The self-referential basis of undecidable dynamics  is fundamentally related to novelty generation~\cite{prokopenko_self-referential_2019}. As pointed out by \citet{markose_novelty_2004,markose_complex_2017}, computational novelty production and ``thinking outside the box'' by digital agents is underpinned by their capacity to encode self-referential statements with negation (e.g., Liar paradox or a G\"odel sentence). Crucially, this capacity allows the agents to exit from known listable sets (e.g., actions, technologies, phenotypes) and produce new structured objects. As discussed earlier (Section \ref{expr-ref}), this is a consequence of the fundamental discrepancy between (the spaces of) expressions and referents which is exploited by various diagonalisation arguments.

\citet{svahn2023ansatz} explored how undecidability which places computational limits on a formal system can manifest itself in biological RNA automata. In particular, they considered Turing-equivalent RNA automata, i.e., the ones that can simulate, and be simulated by, a universal Turing machine. Importantly, they argued that the evolutionary space for these RNA automata can be expanded in a continual process analogous to a hierarchical resolution of computational undecidability by a sequence of Turing’s oracles (and hence, by a sequence of Turing’s ordinal logics and Post’s extensible recursively generated logics). The proposed \textit{ansatz}  hypothesised that the resolution of possible undecidable configurations in biological RNA automata may represent a novelty generation mechanism, in context of  interactions between the automata and their environment~\cite{svahn2023ansatz}.

Thus, computational novelty can be seen as the problem-space expansion, created by agents that use the diagonalisation argument in exploiting the expression-referent discrepancy. In other words, novelty can be generated by \textit{self-modelling agents} that have access to results of meta-level computation --- for example, meta-simulation by Turing oracles and extensible logics (see Sections \ref{open-ended-incomp} and \ref{ordinal}), or receive the corresponding information from external environment.

\subsection{Tangled hierarchies and strange loops}
\label{tangled_hiearchies_def}

In his seminal book, \citet{hofstadter_go_1980} explored how self-reference leads to emergent complexity in systems, by examining recursive structures which appear in a self-similar way at different levels or scales. Importantly, Hofstadter discussed Gödel's Incompleteness theorems, which show that in any formal mathematical system that is expressive enough to describe basic arithmetic, there will be statements that are true but cannot be proven \textit{within the system}. 

\citet{hofstadter_go_1980} proposed and extensively discussed the notion of  \textit{tangled hierarchies} (see Box 5) emerging in a wide range of phenomena, from language and cognition to artificial intelligence and consciousness. Self-reference plays a key role in this process, as it allows for feedback loops and recursive interactions between different levels of the hierarchy. 
In particular, Hofstadter argued that tangled hierarchies are fundamental to human cognition, which relies heavily on analogical thinking and self-referential loops. The concept of the \textit{strange loop} encapsulated this idea, showing that patterns of thought may loop back on themselves to generate \textit{new} levels of understanding.

\subsection{Causative efficacy of biological information}

\citet{walker2013algorithmic} argued that a hierarchy does not just encapsulate some quantity of information, but offers a specific information-processing arrangement describing its organisation, e.g., feedback loops, active information, and integrated information. Such an arrangement gives the higher levels in the hierarchy more functional ``power'', i.e. causal efficacy.  In other words,  information hierarchies may emerge (or self-organise) because they capture the \textit{causative efficacy of information}: 
\begin{quote}
``...biological information has an additional quality which may roughly be called ‘functionality’ --- or ‘contextuality’ --- that sets it apart from a collection of mere bits as characterized by its Shannon information content. ...DNA is a (mostly) passive repository for transcription of stored data into RNA, some (but by no means all) of which goes on to be translated into proteins. ...It is the functionality of the expressed RNAs and proteins --- not the bits --- that is biologically important''~\cite{walker2013algorithmic}. 
\end{quote}

\citet{walker2013algorithmic} further pointed out that biological information (i.e., functionality) is actively involved in  information processing, e.g., control and feedback. Thus, the functionality is dynamic, depending on both the current state and the history of the organism. This perspective --- \textit{the algorithmic origins of life} --- suggests that in order to delineate the phases of non-life and life  one needs to analyse dynamical information and identify causal architecture:
\begin{quote}
``...the emergence of life may correspond to a physical transition associated with a shift in the causal structure, where information gains direct and context-dependent causal efficacy over the matter in which it is instantiated''~\cite{walker2013algorithmic}.
\end{quote}
Following \citet{davies1999fifth}, this work interpreted the phenotype-vs-genotype relationship as hardware-vs-software, identifying chemistry in living systems with hardware,  and information (e.g., genetic and epigenetic) with software: ``the chicken-or-egg problem, as traditionally posed, thus amounts to a debate of whether analogue or digital hardware came first''~\cite{walker2013algorithmic}.
In a related study, \citet{walker2012evolutionary} discussed how the information efficacy and top-down causation can also be applied to the major evolutionary transitions~\cite{szathmary1995major}.

\subsection{Evolution as collective information-processing dynamics}

The emergence of genetic code can be modelled as an information-preservation phenomenon in the presence of noise~\cite{prokopenko_stigmergic_2009}. In particular, the capacity to symbolically represent nucleic acid sequences was argued to emerge, overcoming the ``coding threshold''~\cite{woese2004new}, in response to a change in environmental conditions. In modelling the emergence of universal coding, \citet{prokopenko_stigmergic_2009}  proposed the concept of ``stigmergic gene transfer'', and considered interacting proto-cells as a dynamical system, within which “proto-symbols” encoding features of individual cells are stigmergically shared. 

The model demonstrated that a joint encoding can emerge as a symbolic representation of the shared dynamics, in order to preserve information about attractors of the dynamics in a noisy environment: ``...the pressure to develop a distinctive ``symbolic'' encoding only develops if the noise in the original system is in a particular range, not too small and not too large''~\cite{prokopenko_stigmergic_2009}.

\citet{goldenfeld_life_2011} described evolution  from the standpoint of non-equilibrium statistical mechanics, as a problem in which the key dynamical modes are collective. Taking a provocative stance --- ``Life is Physics'' --- they argued that unifying principles of collective behaviour which arise from physical interactions are applicable to biology, especially in context of the interplay between evolution and environmental fluctuations. 

In exploring the principles underlying collective behaviour, \citet{goldenfeld_life_2011} posed a key question: ``How is it that matter self-organizes into hierarchies that are capable of generating feedback loops which connect multiple levels of organization and are evolvable?''~\cite{goldenfeld_life_2011}. Importantly, the study related co-evolutionary processes to far-from-equilibrium dynamics and pointed out that the corresponding physical laws remain unknown. Nevertheless, their review highlighted a salient connection between co-evolutionary/ecological dynamics and game-theoretic interactions which may generate paradoxical collective outcomes (e.g., Nash equilibria) and even chaotic dynamics (in dynamic game-theoretic models~\cite{sato2002chaos}). 

\subsection{Evolving self-modelling dynamical systems}
\label{self-model}

As discussed in Section \ref{tension},  \citet{watson2022design} studied the “evo-ego” relationship, describing a tension between individual and group interests. While not framing their analysis in terms of ``tangled hierarchies'', the study nevertheless related the organisation of individual self-interested entities to their long-term collective interest, posing two questions:  
\begin{quote}
``(a) what kind of functional relationships between components are needed to make a new individual, and how they need to be organised; and (b) how the organisation of these relationships arises `bottom-up,' i.e., without presupposing the higher-level individual we are trying to explain''~\cite{watson2022design}.
\end{quote}
The study pointed out that the chicken-and-egg dilemma (i.e., a strange loop), encapsulating these two questions, is typical across the major evolutionary transitions in individuality, as it is challenging to answer whether the higher-level unit of selection (needed for complex adaptations) emerges before or after the complex adaptations (needed to generate the higher-level unit of selection). 

In attempting to resolve this conundrum, \citet{watson2022design} drew an analogy with unsupervised \textit{deep learning}, arguing that when individual reinforcement or selection modifies the strength of  inter-unit relationships (analogously to updating the neural network weights during reinforcement learning), ``the system becomes a self-modelling dynamical system'' with predictable system dynamics. In other words, based on interactions with the environment but without an explicit external guidance, a self-modelling system is able to discover some structure in its own constraints and dynamics --- that is, construct its own model. This is analogous to unsupervised learning algorithms which, given the unlabelled input data, can analyse and learn patterns by clustering similar data points or reducing dimensionality.

\subsection{Information integration within collective action}
\label{collective-action}

In formulating the approach of \textit{evolutionary connectionism} based on the analogy with machine learning, \citet{watson2022design} identified 
the conditions required to evolve a novel organisation --- a new level of individuality --- comprising ``short-sighted, self-interested entities'', with the conditions necessary to learn \textit{non-decomposable functions}. Put simply, the relationships between evolutionary units must be organised bottom-up in such a way that their interactions are synergisitic, producing “more than the sum of the parts”. Crucially, this distributed (unsupervised) learning can occur without an explicit system-level feedback, by exploiting regularities encountered at localised interactions. 

In other words, interactions of bottom-level entities produce distributed learning dynamics and integrate information by computing a non-decomposable (i.e., non-linearly separable) function of input states, for example, between some ``embryonic'' collections of particles and the corresponding ``adult'' collective phenotypes~\cite{watson2022design}. 
In turn, this integrated information facilitates collective action, for example, generating ``specific coordinated responses in multiple downstream variables'' --- thus, signifying a downward causation. 
In terms of tangled hierarchies, the bottom-up process, generating a new step in evolutionary individuality, involves synergistic computation of non-decomposable functions which confer collective fitness, thus constraining interactions of the constituent units in a top-down way.

\subsection{Evolutionary connectionism vs ecological scaffolding}
\label{connect-scaffold}

\citet{watson2022design} contrasted their “evolutionary connectionism” approach  with “ecological scaffolding”~\cite{black2020ecological} (see Section~\ref{scaffold}), emphasising that in the latter one has to assume existence of some ``fortuitous extrinsic conditions'' that trigger the population structure to generate specific selective pressures. In other words, they argued that ecological scaffolding resolves the chicken-and-egg problem of the evolutionary transitions in individuality by (temporarily) allowing the environment to assume the role of a “chicken”.  Nevertheless, \citet{watson2022design} pointed out that, even under these conditions, 
\begin{quote}
``...individual selection at the lower level supports the evolution of characters that access synergistic fitness interactions, changing the relationships among the particles, and given that synergistic fitness interactions among particles have evolved, it is subsequently advantageous for particles to evolve traits that actively support this grouped population structure''~\cite{watson2022design}. 
\end{quote}
It can be argued that the ecological scaffolding is another way to shape a tangled hierarchy --- albeit, without \textit{self-modelling}. Once/if the scaffolding becomes redundant, the evolved ``strange loop'' endogenously captures the collective dynamics, comprising both (i) the top-down distribution and dispersal of resources and (ii) the bottom-up computation shaped by the division of labour~\cite{black2020ecological}:
\begin{quote}
``Now the original extrinsic ecological conditions might change or cease, but the population structure necessary to support higher-level selection is nonetheless maintained, supported by the adaptations of the particles''~\cite{watson2022design}. 
\end{quote}
In subsequent analysis, we will argue that developing the self-modelling capacity is a crucial step in biological complexification, improving the organism's capacity to adapt, replicate and function autonomously.

\begin{figure}[!h]
    \centering
\includegraphics[width=1.0\columnwidth]{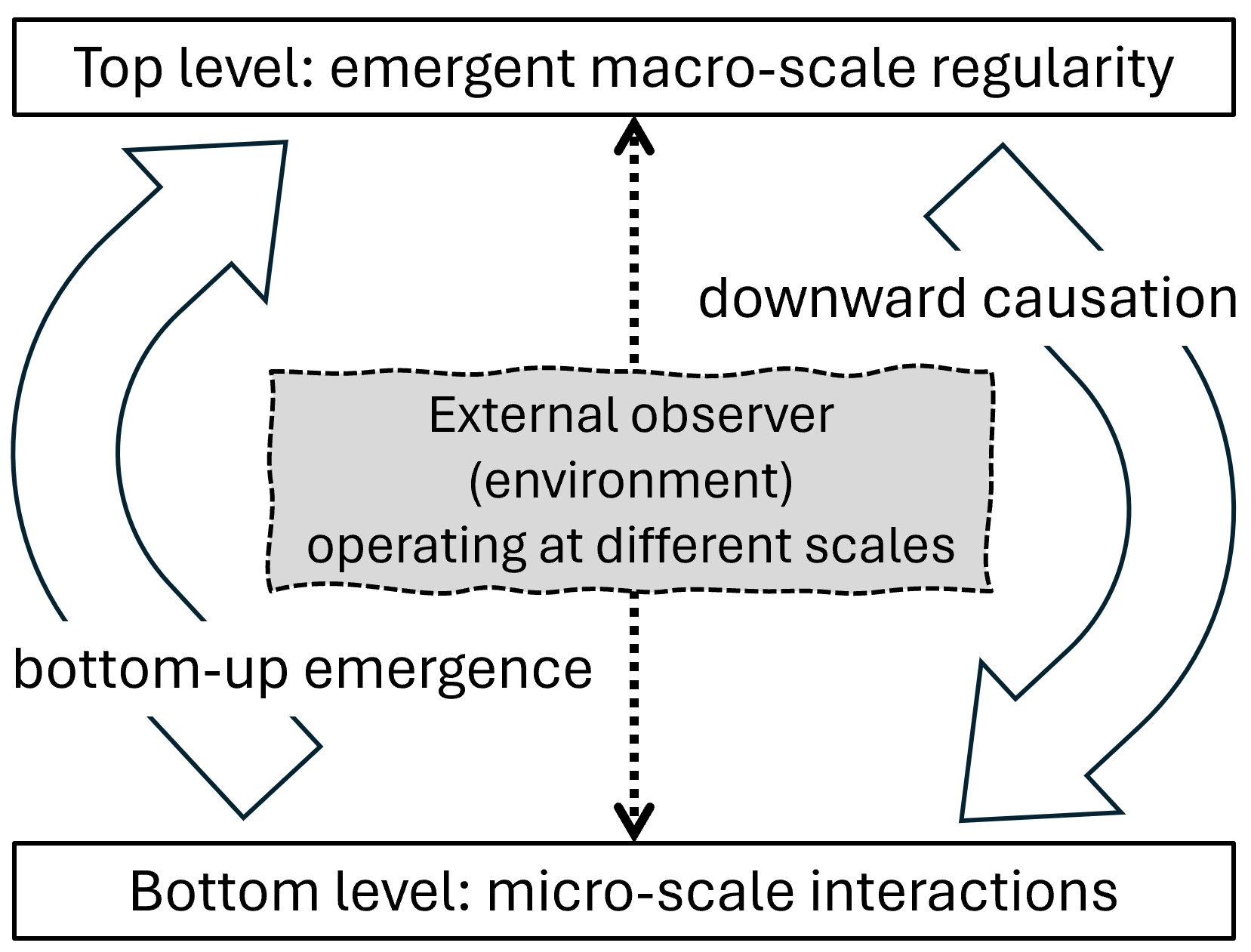}
    \caption{A tangled hierarchy (strange loop) with an external interpreter (e.g., observer, environment, interacting particle, etc.) which may operate and respond to regularities detected at different scales.}
    \label{fig:meta-level}
\end{figure}

\

\begin{figure}[!h]
    \centering
\includegraphics[width=1.0\columnwidth]{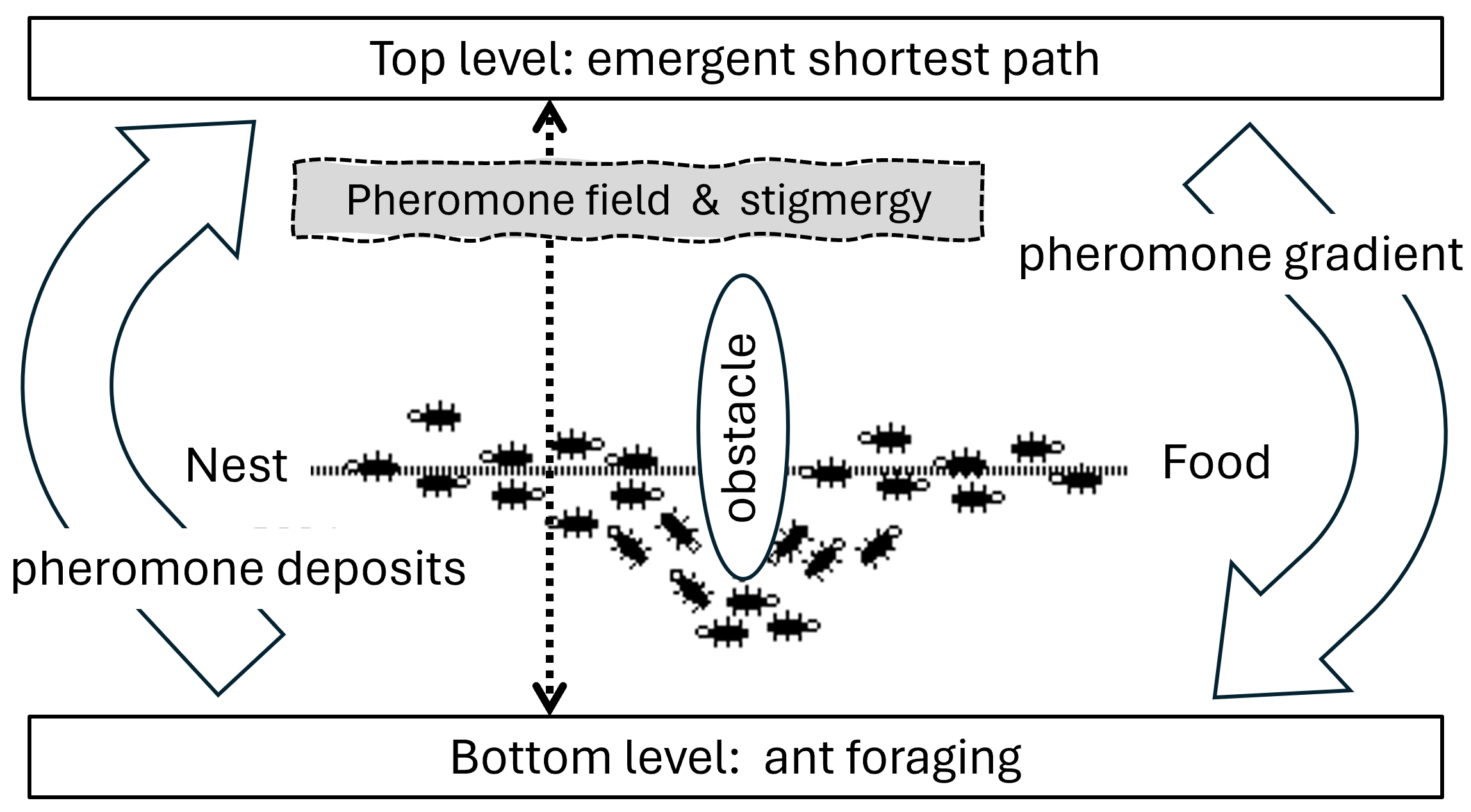}
    \caption{An example of tangled hierarchy (a strange loop): ants foraging for food deposit pheromones which diffuse and evaporate in the environment; ants direct their movement towards locations with higher pheromone concentration,  utilising stigmergy (indirect coordination through the environment); a shortest path emerges out of these interactions (bottom-up emergence), and influences the ant foraging (downward causation). }
    \label{fig:ants-path}
\end{figure}

\boxsection{

{\textbf{Box 5: Self-referential dynamics}}
\label{box:hier}

\subsubsection*{Tangled hierarchies}
Tangled hierarchies are systems in which different levels of organisation or abstraction are intertwined in a way that defies simple hierarchical categorisation: ``an interaction between levels in which the top level reaches back down towards the bottom level and influences it, while at the same time being itself determined by the bottom level''~\cite{hofstadter_go_1980}. In other words, tangled hierarchies interleave bottom-up emergence and top-down (downward) causation~\cite{ellis2012top}. Fig.~\ref{fig:meta-level} illustrates the concept of tangled hierarchy, and Fig.~\ref{fig:ants-path} shows an example: shortest path formation during ant foraging.

\subsubsection*{Strange loops}
As described by \citet{hofstadter_go_1980}, ``a strange loop is a hierarchy of levels, each of which is linked to at least one other by some type of relationship. A strange loop is self-referential, meaning that it loops back on itself in some way. Despite the appearance of movement or progression, a strange loop ultimately returns to its starting point, creating a sense of paradox or contradiction.''  A strange loop involving ant foraging and optimal path formation, shown in Fig.~\ref{fig:ants-path} (see the example described in Section~\ref{ants}), lacks a preferred causal direction: are the ants driven by the pheromone gradient of the path (downward causation), or is the shortest path itself being generated by the ants movement (bottom-up emergence)?
}

\vspace{1cm}

\boxsection{

{\textbf{Box 6: Scales, levels and emergence}}
\label{box:emergence}

\subsubsection*{Scales vs levels}

The difference between scales and levels: ``scales'' imply a (continuous) progression along a single dimension (e.g., from fine-grained to coarse-grained), while ``levels'' typically suggest discrete distinct stages within a hierarchical structure.

\subsubsection*{Meta-levels and interpretation}

There may be an external observer interpreting the interaction between the levels of a (tangled) hierarchy. This external interpreter is not part of the hierarchy, but may exchange matter and information with both interacting levels. For example, the environment may serve as an external, meta-level observer (or meta-simulator), determining the organism's fitness, and thus providing an ``interpretation'' of the organism behaviour/dynamics.  While hierarchies have ``levels'', observer may employ different ``scales'' of observation: high-resolution (e.g., statistical mechanics), and low-resolution (thermodynamics)~\cite{prokopenko2009information}, distinguishing emergent phenomena by detecting and responding to regularities at a particular scale of observation, see Fig.~\ref{fig:meta-level}.
}

\section{Tangled hierarchies and self-modelling}
\label{tangled}

Having examined cross-disciplinary views on novelty generation in biological, computational and physical dynamical systems, we point out a common feature of complexification processes: emergence of hierarchical structures comprising lower-level (e.g., individual) and higher-level (e.g., collective) information-processing elements. These tangled structures exhibit varying degrees of self-reference, self-modelling and self-replication. 

To explore the role played by self-modelling in the emergence of replicating information-processing hierarchies, in this section we propose a distinction between two types of tangled hierarchies: type I (strange loop without self-modelling) and type II (strange loop with self-modelling). We exemplify these two types in several  contexts (subsection \ref{examples-TH}), and argue that different mechanisms are employed to replicate their dynamics (subsection \ref{repl-TH}).

\subsection{Two types of tangled hierarchies}

We define a tangled hierarchy of type I (TH-I) as one where macro-scale information patterns which emerge from the micro-scale interactions within the environment are not compressed, encoded or decoded. In other words, no emergent macro-scale regularity is explicitly modelled in its entirety by micro-scale objects. Despite presence of a recursive flow between levels which ``loops back on itself''~\cite{hofstadter_go_1980}, this loop does not involve compression --- and hence, there is no object in TH-I that is explicitly referred to as `self'. 

In contrast, a tangled hierarchy of type II (TH-II) is one where micro-scale objects operating at the bottom level contain an encoded, compressed representation of some information pattern emerging at the macro-scale. By utilising the compressed information, the micro-scale objects are capable of representing --- that is, modelling --- some of the macro-scale regularities. This modelling is used within a self-referential loop by  alternating encoding and decoding processes. As a result, the compressed representation (i.e., a model) of a higher-level pattern, encoded at a lower level, augments the entire hierarchy with \textit{self-modelling}, due to the tangled nature of the inter-level relationship (see examples of TH-II in subsections~\ref{geno-pheno-ex} and~\ref{evo-ego-ex}). 

In general, there may be a spectrum of hierarchies between types I and II, with some tangled hierarchies utilising compression/modelling to a partial extent, and the general dichotomy we proposed essentially shows the limit cases.

\subsection{Examples of different TH types}
\label{examples-TH}

In order to illustrate the two TH types and build a better intuition, we present several examples, mostly drawn from biology.

\subsubsection{Ant foraging, stigmergy and optimal path formation}
\label{ants}

Ant foraging and optimal path formation in a pheromone field is a tangled hierarchy without self-modelling (TH-I). 
In this case, the tangled hierarchy is constituted by the ants foraging behaviour at the bottom level and the shortest path emerging at the top level (see Fig.~\ref{fig:ants-path}). 
Each ant foraging for food deposits pheromone in response to only local information, without reference to the global (optimal) path. The path itself emerges as a result of stigmergy~\cite{theraulaz1999brief}: indirect interactions and coordination of the ants through the environment (Fig.~\ref{fig:ants-path}). Importantly, ants perceive only local differences in the pheromone field (the pheromone gradient), and make only local field updates. The optimal path is a regularity within the pheromone field --- however, this regularity is not compressed, and the information patterns underpinning stigmergy remain distributed through the environment, without any encoding, decoding or self-modelling.

It can be argued that some features of the emergent path (e.g., the likelihood of it being a shortest path) are attributable to (i.e., partially and indirectly encoded within) the ants genome. In other words, while the ant foraging behaviours belong to the ants immediate phenotype (being directly influenced by their genetic makeup), the class of optimal paths can be seen as a part of the ants extended phenotype~\cite{dawkins2016extended}. That is, the species-dependent pheromone paths, shaped by the interactions between the ants and the environment, contribute to the genome evolution. In this process, the path optimality provides evolutionary rewards resulting in an eventual spread of the beneficial, fitness-increasing genotype. The encoded features are specifically the ones which are likely to produce shortest paths, given the environmental factors. 

However, any actual physical path emerging in a given environment --- a specific regularity in the environment-dependent pheromone field --- is not \textit{directly} encoded in the genome. Rather, it emerges from the behaviours guided by genetic predispositions, while being influenced by the ants interactions with their environment. 

Thus, this TH-I relating specific pheromone paths and ant foraging behaviours is not equipped to make use of a fitness-increasing encoding of some optimal features present in the extended phenotype, as it does not include self-modelling. Our next example, the genotype -- phenotype relationship illustrates a TH-II, in which self-modelling plays a key role.

\subsubsection{Genotype--phenotype relationship}
\label{geno-pheno-ex}

\begin{figure}[!h]
    \centering
\includegraphics[width=1.0\columnwidth]{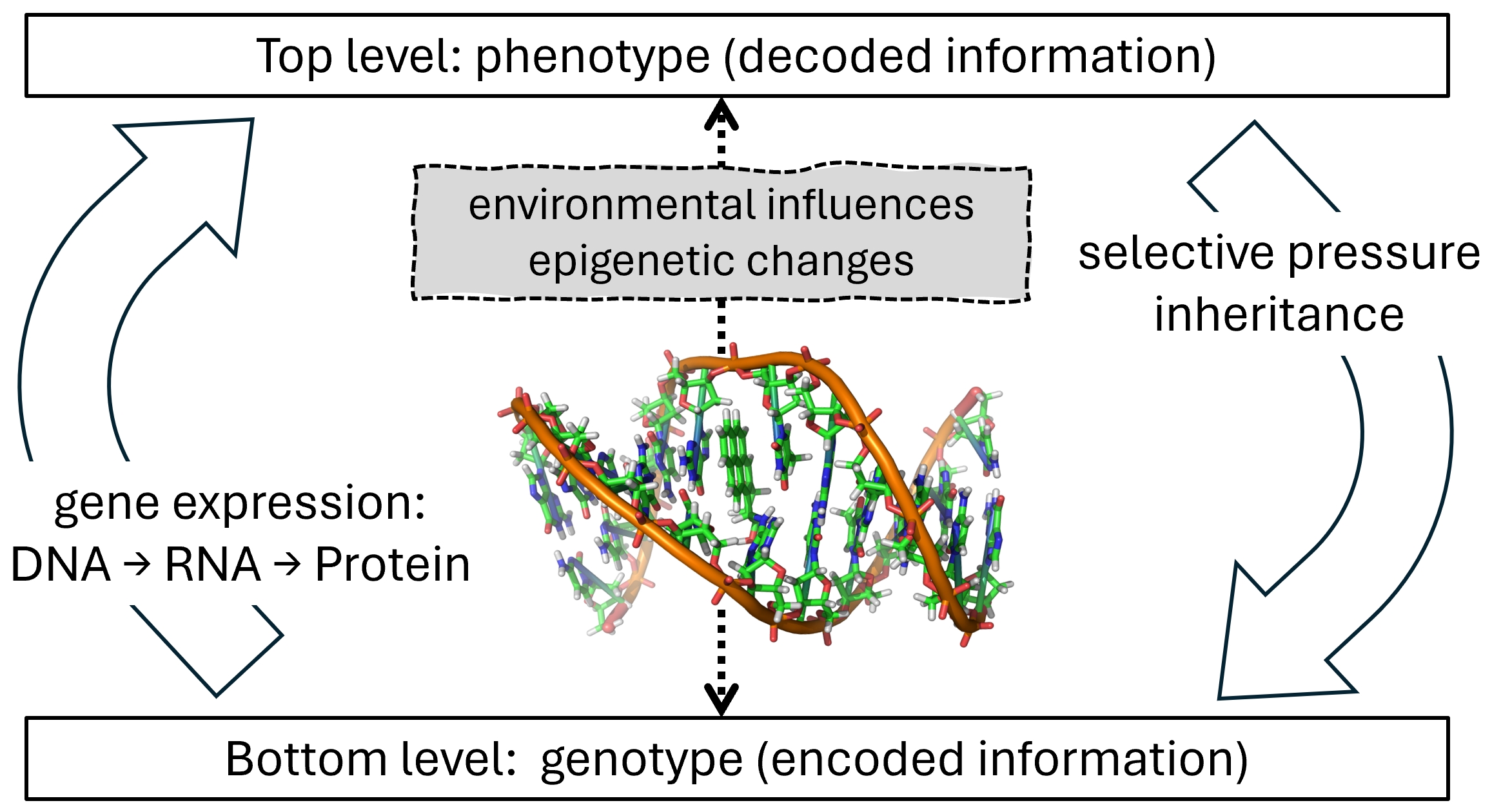}
    \caption{Genotype-phenotype relationship. Gene expression involving transcription and translation of encoded genetic information leads to the decoding of an organism with observable phenotypic traits (bottom-up construction). Given the phenotype, selective pressure and inheritance influence genotype over generations, updating the encoded genetic information which encodes the phenotype itself (top-down causation). Genotype-phenotype relationship is realised in context of environmental influences, epigenetic changes and other factors. DNA image: Wikipedia~\cite{wiki-DNA}.}
    \label{fig:geno-pheno}
\end{figure}

The relationship between genotype and immediate phenotype involves a complex self-referential biological dynamics between the encoded genetic information and the organism itself in presence of environmental influences, as illustrated in Fig.~\ref{fig:geno-pheno}.  Genotype-phenotype relationship can be interpreted as a strange loop between the ``self'' (phenotype) and another object (genotype): the genotype ``encodes'' (models) the phenotype and is also ``decoded'' by the phenotype (see Box 1).  With sexual reproduction, the fitness of a genotype is manifested through its phenotype. In turn, the fitness of a given phenotype, i.e., the organism's chances of survival and reproduction, varies across different selective environments~\cite{kinsler2020fitness}.  Hence, we interpret genotype-phenotype relationship as a tangled hierarchy with self-modelling (TH-II).

\subsubsection{Phyllotactic patterning in plants}

Another example of TH-I is the process of leaf positioning in plants. Leaves typically form at regular intervals at the plant apex, often creating recognisable spiral patterns. These patterns, however, are not directly encoded by the genome, but instead arise due to cell-cell interactions involving a feedback loop. The hormone auxin which triggers leaf outgrowth~\cite{reinhardt2000auxin} is distributed by directional cellular transport~\cite{reinhardt2003regulation}. Each cell assesses the concentration of auxin in neighbouring cells and transports auxin towards those neighbours in proportion to their auxin concentrations~\cite{jonsson2006auxin,bhatia2016auxin}. This feedback, between auxin and its own transport, means that cells that happen to start with high concentrations receive even more auxin from neighbouring regions. However, at a certain distance defined by the relative strength of diffusion, another auxin peak forms and then another, as new space is created through cell divisions~\cite{jonsson2006auxin}. The resultant auxin distribution patterns are not explicitly encoded or compressed, and hence, cannot be replicated without replicating the entire system.

\subsubsection{Evolution of self-modelling collective dynamics}
\label{evo-ego-ex}

The “evo-ego” relationship between the “embryonic” evolutionary units which synergistically interact and integrate information in producing their “adult” collective phenotypes~\cite{watson2022design} is an example of TH-II. As pointed out by \citet{watson2022design}, the relationship between the lower-level (individual) and higher-level (collective) units of selection is a \textit{self-modelling} dynamical system in which integrated information patterns encode (compress) non-decomposable functions of input states (see Section \ref{collective-action}). Thus, we again encounter a self-referential loop with self-modelling: ``the key problem is that evolution is self-referential, i.e. the products of evolution change the parameters of the evolutionary process”~\cite{watson2016evolutionary}.

\subsubsection{Ecological scaffolding without self-modelling}

As noted in Section \ref{connect-scaffold}, a tangled hierarchy comprising collective interests may shape \textit{without self-modelling}, using ecological scaffolding, i.e., information about the evolutionary niche~\cite{black2020ecological,watson2022design}. In this case, there is no self-modelling, as the division of labour does not use any compressed or encoded information about distributed and dispersing resources, and hence, this is an example of TH-I.

\subsubsection{Visual paradoxes}

We now turn our attention to two well-known visual paradoxes, considering whether they may illustrate tangled hierarchies of different types.  

\begin{figure}[!h]
    \centering
\includegraphics[width=1.0\columnwidth]{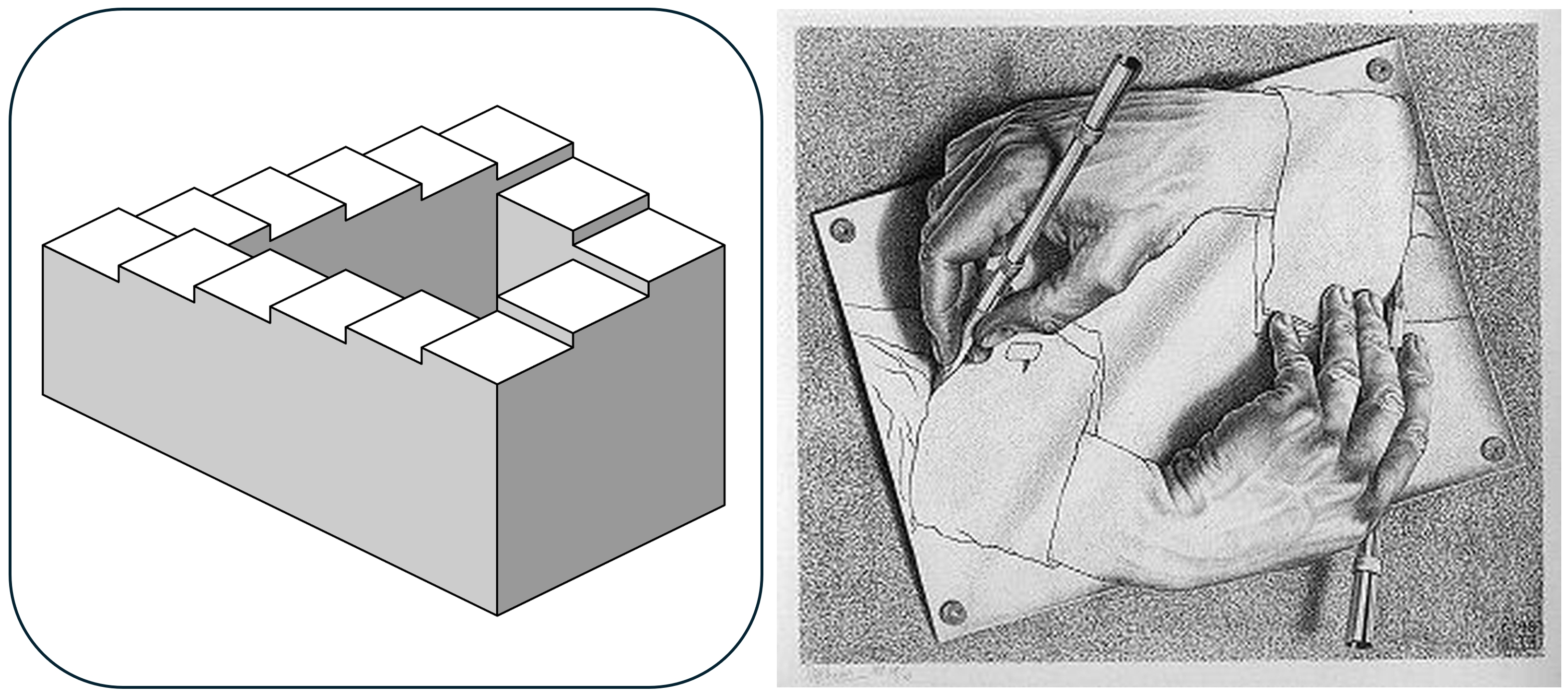}
    \caption{Left: ``Impossible staircase'' (1958) \cite{penrose1958impossible,wiki-Penrose_stairs}. Right: ``Drawing Hands'' (Escher, 1948)~\cite{wiki-Drawing_Hands}.}
    \label{fig:paradoxes}
\end{figure}

The first paradox is an impossible staircase pattern (Fig.~\ref{fig:paradoxes}.Left), devised in 1937 by Oscar Reutersvärd~\cite{torre2019impossible}, rediscovered in 1958 by Lionel Penrose and Roger Penrose \cite{penrose1958impossible}, and artistically implemented in Escher's  lithograph print ``Ascending and Descending'' (1960). In this pattern, the stairs are connected in an impossible way, i.e., the recursion is used within an impossible spatial configuration. Despite the appearance of a continuous staircase, there is an infinite loop of ascent and descent, so that the staircase appears to loop back on \textit{itself}, creating an optical illusion of infinite repetition. However, there is no compression of any information pattern within a self-encoding object, and no self-modelling, suggesting that this is a tangled hierarchy of type I.

In ``Drawing Hands'' (1948), Escher created a visual paradox by depicting two intertwined hands drawing each other into existence (Fig.~\ref{fig:paradoxes}.Right). Each hand is in the process of perpetually drawing the other, creating a recursive loop of drawing an object which draws \textit{itself}. This lithograph not only explicitly captures the idea of self-reference and the infinite recursive loop, but arguably depicts how one object ``models'' (constructs) the other.  

Yet, there is no discernible compression of the model in ``Drawing Hands''   \textit{per se}, and so this example can be called a tangled hierarchy of type II only with a stretch. Nevertheless, the notion of ``self'' is  pronounced in this lithograph much more than in ``Impossible staircase'' and ``Ascending and Descending'', showing a transient stage towards fully-formed TH-II, in which  self-encoding may be achieved by compression.  In short, ``Drawing Hands'' lithograph may illustrate a tangled hierarchy positioned between the extremes of TH-I and TH-II.

\subsection{Replication in tangled hierarchies}
\label{repl-TH}

The distinction between the two types is also evident in different replication mechanisms available to dynamics within TH-I and TH-II. 

In order to replicate information patterns and regularities emerging within TH-I (e.g., a pheromone path), the entire environmental state or complete dynamics needs to be reproduced (e.g., the entire pheromone field or the complete history of ant interactions from the initial state). Without compression and encoding, the information-processing is distributed throughout the entire system (e.g., stigmergy), and hence, needs to be replicated in every detail. 

Computation-theoretically, in order to simulate dynamics unfolding within TH-I, one would require a \textit{trivial} universal simulator which is defined as a simulator describing the collection of all particular solutions --- in other words, it ``needs access'' to all corresponding Turing machines computing possible dynamic trajectories~\cite{gonda2023framework}. In this case, one may recall an insight of  \citet{brooks1990elephants}: ``the world is its own best model --- always exactly up to date and complete in every detail''.

The impossibility to replicate an optimal pheromone path emerging during ant foraging without reproducing the entire pheromone field (including the full pheromone gradient) reinforces our classification of this case as TH-I. If an optimal pheromone path was symbolically (i.e., digitally) encoded somewhere in the system, then it would have been possible to replicate just this encoding and expect that ants would access, ``read-out'' the encoded information, and rapidly rediscover the optimal connectivity. 

An interesting but fatal phenomenon --- ``ant mills'' --- is observed when an environmental effect causes ants to lose the pheromone path and start following each other instead, potentially driving them in ``death spirals'' to exhaustion and death~\cite{schneirla1944unique}. As pointed out by \citet{delsuc2003army}, ``the occasional but deadly formation of circular mills seems to be the evolutionary price that army ants pay to maintain such an ecologically successful and stable strategy of collective foraging''.  Specifically, this pathological behaviour can be explained by the ability of army ants to collectively select a raid direction~\cite{couzin2003self}. Arguably, this behaviour arises when the tangled hierarchy lacks adequate means to explicitly encode and replicate additional information (e.g., a preferred raid direction). In effect, in the absence of an explicit pattern to follow, ants are reduced to follow the only regularity that remains (a circular path). 

In contrast, information patterns and regularities emerging within TH-II can be replicated more efficiently. For example, to replicate a phenotype sufficiently well, one would need to copy/clone a genotype, and place it in a similar environment. This is ensured by the compression and encoding of relevant genetic information within DNA. 
Computationally, to replicate dynamics of TH-II, one needs a \textit{non-trivial ``singleton''} universal simulator which provides a universal solution via a single universal Turing machine~\cite{gonda2023framework}. This solution is more efficient than the trivial universal simulator needed to describe the collection of all trajectories generated by TH-I dynamics.

\section{Emergence of self-modelling in tangled hierarchies}
\label{self-modelling}

In principle, patterns emerging in TH-I (e.g., pheromone paths, hormone distribution gradients, ecological scaffolding) are  compressible due to the presence of regularities. Once such compressed information becomes available to the system, it can then begin to process this information about \textit{itself} --- hence becoming a tangled hierarchy of type II. In this section we investigate the functional advantages enabled by such compression (i.e., the dimensionality reduction), which allows a self-referential system to be more predictive about its dynamics, exploiting the regularities more efficiently.


\subsection{Emergence of functional self-descriptions}
\label{self-descr}

Having discussed the difference between tangled hierarchies of two types, we now consider the questions of how and why a functional self-description may emerge, and what functional advantages it could provide to TH-II relative to TH-I.  It is worth pointing out that TH-I which operates without self-modelling still can maintain a good level of stability and collective efficiency (e.g., ant colonies and other animal groups). And so one needs to examine what additional selective advantages would be offered by formation/emergence of self-modelling.

As discussed by~\citet{mcmillen2024collective}, the higher levels of organisation bring distinct advantages:
\begin{itemize}
    \item a different energy landscape that can be exploited by the collective;
    \item propagation of information across scales;
    \item a new problem-space and “extended patterns of information” available to the collective.
\end{itemize}
In short, the higher levels of organisation can favourably distort the energy landscape for their subunits and thus, provide a ``causal architecture'' enabling synergistic problem-solving competencies. These new competencies allow the collective ``to navigate spaces of which the subunits are unaware''~\cite{mcmillen2024collective}.  This causal architecture can be interpreted as an information hierarchy, tangling individual and collective dynamics. 

Once ``extended patterns of information'' become available within a higher-level problem-space, a functional self-description may emerge to encapsulate the newly accessible regularities. Such self-modelling allows the information hierarchy to compress, encode and preserve beneficial information patterns in presence of stochastic environmental effects. The compressed self-description can then be decoded to recover the function disrupted by the adverse conditions. 

For example, imprecise copying without an encoding was common during the evolutionary period dominated by lateral movement of genetic material via HGT.
\citet{woese2004new} referred to results of such imprecise translation mechanism as ``statistical proteins, proteins whose sequences are only approximate translations of their respective genes'', while a consensus sequence for the various imprecise translations might closely approximate an exact translation of that gene.
In contrast, precise copying using encoded/compressed information is a key feature of VGT, the vertical transmission of genes from parent to offspring.
We can assume that during the period when HGT was the dominant form of gene transfer, the tangled hierarchies of type I could replicate only approximately, by imprecisely copying the whole organism. 

As mentioned in Section~\ref{threshold}, ``coding threshold” has possibly separated the earlier evolutionary stage and the RNA world, providing the capacity to represent nucleic acid life symbolically, in terms of amino acid sequences~\cite{woese2004new}.  Once some proto-codes emerged, encapsulating a prototypical self-descriptions of cellular dynamics, the Darwinian vertical descent based on VGT became the predominant mode of replication. Our conjecture is that this led to formation of the tangled hierarchies of type II. 

Eventually, when the early proto-codes became interchangeable across the VGT replicators, a universal DNA-based code emerged, and the genotype-phenotype relationship fully formed. In other words, \textit{replication robustness} may benefit from a universal code: it is more efficient to make multiple copies if they are described in a universal way, and this provides a selective pressure for the code universality. 

Put simply, a functional self-description enables a more efficient replication process, preserving extended information patterns. Therefore, a compressed information about salient non-decomposable regularities --- a ``model'' of the replicator --- becomes codified in some symbolic (i.e., digital) form. Thus, information preservation within tangled hierarchies of type II which utilise self-modelling improves their structure and long-term function.

There are several possible mechanisms for emergence of functional self-descriptions within tangled information hierarchies. It could be triggered by a division of labour~\cite{ispolatov2012division} which facilitates a split of competencies (i.e., symmetry breaking), followed by embedding or encapsulating some of these competencies within a proto-code. Alternatively, a perturbed biological system may be able to preserve information by finding suitable biochemical elements in its environmental locality and entrapping them as proto-code representing relevant features of its dynamics (e.g., attractors)~
\cite{prokopenko_stigmergic_2009}. As pointed out by \citet{arthur1993evolution}, system's complexity may grow by capturing ``software'', that is, by capturing simpler elements and learning to ``program'' these as ``software'' in order to achieve its own goals.

In summary, functional self-descriptions and self-model\-ling emerge in response to some selective pressures and bring specific evolutionary advantages: robustness of replication, division of labour, preservation of non-decomposable collective dynamics, etc. The general motif is that self-referential and self-modelling dynamics improve efficiency of information-processing within a tangled information hierarchy (TH-II), at the cost of encoding and decoding of a compressed self-representation.

\subsection{Synergistic fitness interactions exploit the discrepancy between ``referents'' and ``expressions''}
\label{discrepancy-fitness}

Our conjecture is that the evolutionary tension between individual and group interests~\cite{watson2016evolutionary, watson2016can,watson2022design,mcmillen2024collective} is an example of a principled and generic ``inconsistency'', directly related to the expression-referent discrepancy discussed in Sections \ref{expr-ref} and \ref{comp-novelty-gen}. Thus, we argue that the major evolutionary transitions in individuality exploited a divergence between the sizes of the expression and referent phase-spaces (i.e., problem-spaces). 

We argue that the tangled hierarchy linking the individual and group levels exhibits the expression-referent discrepancy --- the reason is that not all collective interests are reducible to a (sum of) individual preferences. This creates inconsistencies between individual and group interests (``individual-level selection will oppose the creation and maintenance of adaptations that enforce selection at the group level''~\cite{watson2016can}). Computation-theoretically, our conjecture is that while individual interests are encodable via a given set of referents, the non-decomposable synergistic group interests are not necessarily expressible via the given referents.  
Thus, a disagreement between the individual and group interests can lead to either of two cases:
\begin{itemize}
\item[(i)] in the case when “maximising the utility of the components rather than the collective''~\cite{watson2016evolutionary} dominates, then robustness and stability of the current setup is preferred (there is no transition); 
\item[(ii)] otherwise, the tension is resolved by a transition where ``coordinated phenotypic differentiation is then favoured and can thereby maximise collective fitness''~\cite{watson2016evolutionary}, which in turn expands the phase-space.
\end{itemize}

\section{Biological arrow of time as open-ended meta-simulation}
\label{arrow}

In section~\ref{tangled} we introduced the distinction between tangled hierarchies with and without self-modelling. Then, in section~\ref{self-modelling} we argued that a functional self-description helps the TH-II system to innovate in two ways: by capturing and encoding (i.e., compressing) beneficial regularities and patterns in its dynamics, and preserving the resultant extended information patterns. This, in turn, facilitates a more efficient replication process.

In this section, we build on this premise and formulate our central conjecture, hypothesising that tangled hierarchies that emerge at various stages during biological evolution develop and expand in a continual, open-ended, process of self-referential dynamics and meta-simulation. This process encounters and then resolves undecidability by expanding the computational problem-space (``jumping out of the system'') during transitions,  proceeding according to the following steps:

\begin{enumerate}
\item \textit{Emergence of tangled hierarchy without self-model\-ling.} The emergent TH-I (see Sec.~\ref{tangled}) enables distributed coordination (e.g., pheromone paths emerging as a result of ant foraging and stigmergy) and imprecise replication (e.g., ``statistical proteins''), but offers limited stability.
\item \textit{Encapsulation of compressed functional self-descrip\-tion.} This transitional step identifies some regularity, i.e., information pattern, which brings evolutionary advantages (e.g., replication robustness), and then encodes this pattern within the lower-level of TH-I, providing it with a functional self-description. In general, this can be driven by a combination of exogenous and endogenous factors such as symmetry breaking (e.g., division of labour) and information preservation in presence of external noise. 
\item \textit{Formation of tangled hierarchy with self-modelling.} The formed TH-II (see Sec.~\ref{tangled}) is capable of using encoded and decoded self-descriptions (e.g., a fully formed genotype-phenotype map). This has the following consequences:  
\begin{itemize}
    \item \textit{Self-reference.} Given the encoding and decoding capabilities, the formed expression-referent relationship exhibits duality, with expressions potentially becoming referents and vice versa, e.g., program-data duality where programs can be encoded as data and used as inputs (see~\ref{App:TM}).
    \item \textit{Undecidability.} The expression-referent duality inevitably brings undecidability for a given system due to the expression-referent discrepancy: there are always more expressions (objects) than referents (encodings), and some expressions (possible objects) are inaccessible by the system --- essentially, this is a diagonalisation argument. In the context of genotype-phenotype relationship, there are always more possible phenotypes than the genotypes available under a particular encoding scheme.
\end{itemize}  
\item \textit{Extension of the problem-space}, due to the expression-referent discrepancy. In a biological context, this enables organisms interacting with their environment (given their current niche), to access and exploit a new problem-space and “extended patterns of information”~\cite{mcmillen2024collective}, i.e., to perform meta-simulation and resolve current inconsistencies (mismatches) by better fitting the environment~\cite{svahn2023ansatz}. 
\end{enumerate}
In general, the extensions described in step (4) are not meant to be abrupt (notwithstanding the computation-theoretic analogy of Turing jumps), and may develop by a slow accumulation of lower-level changes within subunits adapting over time, until a tipping point. 

The entire process involves bottom-up emergence, self-modelling, non-decomposable collective fitness (i.e., information integration)~\cite{watson2022design}, and “information self-creation”~\cite{heng2023karyotype}. These key features shape the tangled information hierarchies in a way that generates computational novelty. Informally, at a major transition, the collective dynamics may need different “symbols” (a new “code”) to efficiently describe non-decomposable dynamics, relative to the symbols currently available to the individual subunits.

In general, this process follows a directional spiral, returning after step (4) to step (1) and forming a new TH-I, initially without self-modelling. Resolving a tension expands the problem-space which allows the extended system to access a new, more complex, landscape of novel (collective) possibilities. However, the new landscape brings about a new discrepancy (e.g., new biological contradiction, that is, undecidability), thus continuing the process in an open-ended irreversible way, represented by the biological arrow of time. 

We emphasise that the open-ended computational meta-simulation, described in subsection \ref{ordinal}, provides a computa\-tion-theoretic interpretation --- a kind of semantics --- rather than a specific mechanism employing Turing oracles and Turing jumps. In biological systems, oracles do not have to exist \textit{per se} --- instead, higher-level tangled hierarchies emerge by capturing synergistic information patterns within the expanded phase-spaces. In other words, once a discrepancy between expressions and referents is resolved at a given level, the oracle's job is already completed. In summary, the computation-theoretic interpretation of open-ended meta-simulation offers a unifying perspective on diverse real-world dynamics and specific biological processes that exhibit temporal asymmetry. 

\vspace{-0.3cm}

\section{Discussion and conclusion}
\label{discussion}

In this study we we interpreted the phenomenon of open-ended biological complexity as a dynamic computational process and proposed a computation-theoretic argument for the biological arrow of time. Our argument follows G\"odel--Turing--Post recursion-theoretic framework which formalises the construction of extensible computational systems such as Turing $\alpha$-oracle machines. We proposed that this open-ended generation of computational novelty involves meta-simulation performed by higher-order systems that successively simulate the computation carried out by lower-order systems. Essentially, this open-ended meta-simulation provides solutions to the undecidable problems encountered by the lower-order ``subunits'' and hence, expands the effective phase-space of possibilities.  

Before concluding, here we briefly reflect on several studies which proposed fundamental principles for open-ended evolution and biological complexification, and point out connections with our approach.  

\subsection{Evolutionary role of expanded genomes}

\citet{heng2023karyotype} discussed (species-specific) karyotype or chromosomal coding, arguing that
\begin{quote}
``If a change in karyotype coding generates new information at the system level, then an altered karyotype contains the necessary information for the emergence of a new genome system -- the necessary information for macroevolution“~\cite{heng2023karyotype}.
\end{quote}
While some changes in karyotype may promote macroevolution, larger genome or karyotype changes are unlikely to be sufficient for it on their own. Nevertheless, these changes may provide a pre-condition: a large random jump in the fitness landscape, followed by a period where the organism is able to explore locally within the landscape more easily than before. 

The evolutionary role of an expanded genome is also emphasised by \citet{bingham2024nonadaptive} who reported an association between eukaryotic genome duplication and the evolution of multicellularity. This was contrasted with ancestral prokaryotes which tended to lose rather than accumulate DNA, and this may have prevented a transition of prokaryotes to multicellularity. We may interpret this difference between eukaryotic and prokaryotic genomes as the difference in their abilities to form a larger phase-space by adding more ``letters” to their encoding schemes --- essentially, as the difference in their \textit{self-modelling} abilities. 

Interestingly, \citet{bingham2024nonadaptive} also highlighted that eukaryotic cells have a well-developed ability to deal with parasitic elements which form a significant part of eukaryotic genomes. This ability exemplifies an inconsistency-resolving element which may have also facilitated a transition to  multicellularity.  

One may draw a parallel between parasitic elements (invaders) and the ``contrarian” agents (i.e., Liar agents, in the sense of Liar paradox) which   exploit the gaps in self-descriptions, e.g., computer viruses that change the host code to generate the outcome opposite to the intended computation~\cite{marion2012turing}, or novel antigens disrupting the ``self vs non-self” distinction within autoimmune system and causing autoimmune diseases~\cite{geenen2021thymus}. \citet{markose2021genomic} insightfully pointed out that the knockout of specific auto-immune regulators leads to the loss of self-representation for certain self-gene codes within the autoimmune system, thus generating autoimmune pathologies. These examples illustrate how the ability to implement negation contributes to generation of inconsistencies within a tangled information hierarchy with self-modelling.

\subsection{Increasing ``dynamic kinetic stability''}

In an attempt to reformulate Darwinian theory of evolution and extend it to inanimate matter, \citet{pross2011toward} introduced the principle of Dynamic Kinetic Stability (DKS). The DKS principle, expressed in physicochemical terms, favours stable kinetic \textit{patterns} balancing the rates of replication and decay (e.g., during autocatalytic reactions). This approach emphasised that certain relatively simple self-replicating systems (e.g., single molecules, oligomeric sequences comprising more than one subunit, minimal molecular networks) may have used imperfect replication while evolving toward replicating systems of greater DKS. Crucially, \textit{autocatalysis} and \textit{cooperative behaviours} were identified as common features of both abiogenesis (i.e., life emerging from nonliving matter) and biological evolution:
\begin{quote}
    ``...cooperative behavior can emerge and manifest itself at the molecular level, that the drive toward more complex replicating systems appears to underlie chemical, and not just biological, replicators.\\
    ...life’s emergence began with the chance appearance of some relatively simple replicating chemical system, which then began the long road toward increasingly complex replicating entities.''~\cite{pross2011toward}. 
\end{quote}
These elements --- higher-order regularities (i.e., collective or cooperative behaviours) emerging out of interactions of imperfectly replicating subunits; self-referential (autocatalytic) reactions; and the stability of replication dynamics --- can also be interpreted in terms of tangled hierarchies which continually expand their problem-spaces. 

\subsection{Increasing ``functional information''}

\citet{wong2023roles} studied the roles of function and selection in evolving systems, and identified three universal evolutionary mechanisms utilising information about the system--environment interactions: static persistence, dynamic persistence, and novelty generation. Information was interpreted as ``patterns of data in a system that encodes about itself, its environment, or about its relation with its environment'', while functions were interpreted as processes that have ``causal efficacy over the internal state of a system or its external environment''~\cite{wong2023roles}. The study proposed a general law of increasing \textit{functional information}: 
\begin{quote}
``The functional information of a system will increase (i.e., the system will evolve) if many different configurations of the system undergo selection for one or more functions''~\cite{wong2023roles}.
\end{quote}
Among core functions capable of perpetuating themselves, such as  dissipation and homeostasis, this approach also highlighted that self-replicating systems, including living systems, are necessarily autocatalytic, and drew an analogy with the DKS framework. Going beyond the DKS approach, \citet{wong2023roles} identified \textit{information-processing} as another self-perpetuating core function. The described information-centric account distinguished different kinds of dynamical persistence with respect to the distinct levels at which information patterns contribute to persistence. In particular, the study considered (i) information storage: ``memory'' which allows for \textit{encoding} associations, (ii) information inference of future states based on encoded memory: ``memory-based prediction'' which provides a \textit{causal model} and improves persistence, and (iii) counterfactual reasoning: ``prediction outside of memory'' which generates \textit{novelty} through imagining  previously nonexistent versions of reality~\cite{wong2023roles}. 

The increasing dynamic kinetic stability and the increasing functional information reflect the arrow of time. We contend that both these principles can be subsumed by the computation-theoretic G\"odel--Turing--Post characterisation of the open-ended meta-simulation by systems that expand their problem-spaces in the search of ways to resolve specific contradictions and tensions. In particular, ``counterfactuals'' which are required to generate novel functional information~\cite{wong2023roles} can be formed only by considering negation, typically in context of the Liar paradox.  

\subsection{Assembly theory and the ``adjacent possible''}

A recent proposal on ``assembly theory'' (AT) also attempted to explain and quantify selection and evolution in the context of novelty generation:
\begin{quote}
``This approach enables us to incorporate novelty generation and selection into the physics of complex objects. It explains how these objects can be characterized through a forward dynamical process considering their assembly. By reimagining the concept of matter within assembly spaces, AT provides a powerful interface between physics and biology. It discloses a new aspect of physics emerging at the chemical scale, whereby history and causal contingency influence what exists''~\cite{sharma2023assembly}
\end{quote}

In assembly theory, an object is defined through its possible formation histories in an ``assembly space'', so that ``objects are made by joining elementary building blocks together recursively to form new structures''~\cite{ellis2023foundational}. Essentially, given the  building blocks available at the time, the assembly space is a problem-space that comprises possible pathways for assembling an object. 

This view can be compared with the concept of the ``adjacent possible'' proposed by \citet{kauffman2014prolegomenon}: the set of all potential configurations that are just one step away from the current state of a system. These possibilities are constrained by the existing components, structures, or knowledge of the system. Thus, the innovations arise from the existing system's state and the adjacent possibilities that are immediately accessible from that state.

The assembly theory and the ``adjacent possible'' concept identified a discrepancy between the space of objects that can be constructed using actually accessible blocks, and the space of objects that are conceivable. Computation-theoretically, this discrepancy can be interpreted  as the expression-referent discrepancy which generates contradictions, and forces an expansion of the problem-space by meta-simulating and discovering new (assembly) pathways.

\subsection{Social dynamics and undecidability}

It is likely that social complexity increases in an open-ended way as well, along a socio-biological time arrow. However, the social complexification is out of scope for this study. We briefly note, following other similar hypotheses~\cite{arthur1993evolution,woese2004new,goldenfeld_life_2011,mcmillen2024collective,markose_complex_2017}, that the emergence of tangled hierarchies with functional self-descriptions and self-modelling may drive novelty generation in the evolution of language and culture. 

In particular, one may conjecture that grammar provides a functional self-description of natural language, codifying various linguistic elements and rules, as well as their interpretation. 
Similarly, we may interpret social institutions, such as traditions, norms, conventions, laws, legal frameworks,  as functional self-descriptions of society. These social institutions encode and encompass the organised and established patterns of human behaviour and relationships. 

The grammars and institutions which encode the prevailing regularities are always short of describing all possible scenarios which may unfold due to interactions with external environment. These linguistic and social  discrepancies between ``expressions'' (linguistic or social traits) and ``referents'' (codified grammatical or institutional rules) inevitably lead to mismatches, inconsistencies and contradictions. In turn, resolution of these tensions necessitates a construction of expanded problem-spaces, augmented with new conventions (new ``axioms'').

\subsection{Summary}

We used concepts and methods developed within the G\"odel--Turing--Post framework --- such as the expression-referent discrepancy, self-referential dynamics, and Turing jumps resolving undecidability ---  to draw parallels with the tension between individual and collective dynamics, self-modelling processes producing non-decomposable collective fitness, and coding thresholds separating evolutionary transitions. 
We identified \textit{open-ended meta-simulation} as a common thread across all these phenomena, where lower-order systems find ways to incorporate environmental (external) feedback in their self-modelling dynamics, thus supporting ``information self-creation''~\cite{heng2023karyotype} and ``information integration''~\cite{tononi2004information}.

In developing this analogy, we proposed a distinction between two types of tangled hierarchies, with the first one (TH-I) having no self-modelling capabilities, despite involving strange (self-referential) loops (see Section~\ref{tangled}). Under some symmetry-breaking conditions (e.g., division of labour or environmental noise), it becomes beneficial for the system to encapsulate (embed or entrap, see Section~\ref{self-descr}) additional components that encode a compressed representation of salient emergent regularities and information patterns. The tangled hierarchy augmented with functional self-description (TH-II) is capable of self-modelling.

We investigated the biological relevance of the self-modelling capabilities enabled by TH-II systems. On the one hand, self-modelling brings  benefits by allowing the system to explore alternative (adaptive) scenarios, improve its energy and replication efficiencies, as well as the resilience to perturbations. On the other hand, however, self-modelling generates a discrepancy between the space of possible referents (e.g., genotypes) and the space of possible expressions (e.g., phenotypes). Indeed, while the space of objects that can be expressed by the available code is limited, the space of possible objects is necessarily larger (see Section~\ref{discrepancy-fitness}). This discrepancy is the core element exploited by diagonalisation arguments which construct at least one such ``extra” object, demonstrating undecidability. Thus, self-reference plays a role in formulating inconsistencies in style of the Liar paradox by exploiting the expression-referent duality (e.g., program-data duality), when an encoded expression (e.g., program) is recursively used as a referent (e.g., data input) for itself (see Section~\ref{comp-view} and \ref{App:header}). 

Having related computational and biological undecidability, we argued that meta-simulation by the organism-environment interactions is expected to resolve these inconsistencies (i.e., mismatches and contradictions framed as the Liar paradox). This resolution generates computational and/or biological novelty, by encapsulating the identified contradictions as new axioms or adaptations, thus extending the system and forming a new phase-space (Section~\ref{arrow}). 
By analogy with computational undecidability, we point out that the process of open-ended meta-simulation requires (i) the expression-referent duality which forms within TH-II with self-modelling, (ii) an infinite computational medium, and (iii) the ability to implement negation. 

Finally, we suggested that the biological arrow of time generates novelty by expanding problem-space via ``Turing jumps” (not necessarily abrupt) which construct oracle machines (meta-simulators) of higher order (Section~\ref{arrow}). This argument adds ``information self-creation” to the three canonical elements of computation: information preservation (memory/storage), information modification (processing), and information usage (communication or transmission)~\cite{langton1990computation,prohaska2010innovation,lizier2012coherent,kershenbaum2016acoustic,heng2023karyotype}. We propose that, in terms of dynamical systems, it is precisely the information self-creation that forms a dimension for major evolutionary transitions along the biological arrow of time.

\appendix

\section{Self-reference and undecidability}
\label{App:header}

\subsection{Models of computation: Turing Machines}
\label{App:TM}

A Turing machine is formally defined as a system comprising a set of states, with several states designated specifically (the \textit{start} state; and two halting states \textit{accept}  and \textit{reject} --- or alternatively, only one halting state $halt$), the input and tape alphabets, and the transition function that maps pairs of a state and a tape symbol to  triples of a state, a tape symbol and a tape shift (left or right)~\cite{Sipser,hopcroft1969formal}.

Using the definition with two halting states, denoted ($\ddagger$), the output of halting computation is the decision whether the input (initially contained on the tape) is accepted or rejected. Specifically, if the machine halts in the accept $accept$ state, then the initial tape contents (i.e., the input) is considered to be accepted, and if it halts in the $reject$ state, the input is considered as rejected.

Equivalently, one may define a Turing machine with only one halting state $halt$, so that if the machine halts by reaching this state, then the content written on the tape at halting time represents the actual computation output; this definition is denoted ($\dagger$). 

The flexibility of representing the computation target, provided by these two definitions, underscores the duality of the program and the data: a part of the input data may always be used within the computing machinery, and vice versa~\cite{prokopenko_self-referential_2019}.

\subsubsection{Halting problem and diagonalisation}
\label{diag}

Computational undecidability can be demonstrated by the interplay between a hypothetical \textit{universal decider}, logical negation and self-referential encodings. 
Here, we outline a proof of the undecidability of the Halting problem \cite{Sipser}, i.e., considering whether there exists a computer program $D$  that  can determine, for any program $M$ and input $I$, whether $M$ terminates on $I$. In other words, if it existed, such a universal decider $D$ would terminate for all pairs $(M, I)$. Turing proved that no such program $D$ exists for Turing Machines~\cite{turing_halting_1937}. 

Using the diagonalisation argument (see the expression-referent grid (\ref{diagonalization}) in Section~\ref{diag-prelim}), we can take expressions $E_i$ to be programs, referents $\NGodel{E_i}$ to be the encoding (e.g., source code) of the corresponding program (as an input string), and properties $P_{ij}$ taking values \textit{accept}/\textit{reject} if program $E_i$ terminates/runs-forever on input $\NGodel{E_j}$. 

We begin by assuming that the Halting problem is decidable, i.e., $P_{ij}$ can always be determined. We then continue by constructing a ``Liar program'' $E_v$ whose outputs for each input $\NGodel{E_j}$ is defined by $\neg P_{jj}$, where $\neg$ flips \textit{accept} and \textit{reject}. In other words, outputs $P_{vj}$ are defined by inverting the corresponding diagonal entries $P_{jj}$. The contradiction of the decidability of the Halting problem follows when one considers the value of the diagonal entry $P_{vv}$ which can not evaluate to either \textit{accept} or \textit{reject}.

\subsubsection{Self-referential Liar machine}
\label{inverter_liar}

In describing an alternative way to produce an equivalent contradiction we follow \citet{Sipser} (see also \citet{prokopenko_self-referential_2019} for comparative analysis of this argument across Turing machines, formal systems and cellular automata). 

\textbf{Step 1.} We assume that there exists \textit{a decider} machine $D$ such that on input $[M, I]$, where $M$ is a TM and $I$ is an input string, the decider $D$ halts and accepts its input if $M$ accepts $I$, and otherwise, if $M$ fails to accept $I$, $D$ halts and rejects its input:
\begin{equation}
\label{t:p}
    D([M, I]) = 
\begin{cases}
accept \hspace{7mm} \textrm{if} \ M \ \textrm{accepts} \ I    \\
reject \hspace{7mm}  \textrm{if} \ M \ \textrm{does not accept} \ I                     
\end{cases}
\end{equation}

\textbf{Step 2.} We construct \textit{the inverter (``Liar'')} machine $L$ that 
\begin{enumerate}[(i)]
    \item  interprets its input $[M]$ as  encoding of some TM $M$,
    \item invokes the decider machine $D$ with compound input $[M, [M]]$,
    \item inverts the outcome of $D$.
\end{enumerate}
In other words, the inverter (``Liar'') machine $L$ accepts the input $[M]$ if $D([M, [M]])$ rejects its compound input (i.e., by definition of $D$, if $M$ does not accept $[M]$), and  rejects the input $[M]$ if $D([M, [M]])$ accepts its compound input (i.e., if $M$ accepts $[M]$):
\begin{equation}
\label{t:d}
    L([M]) = 
\begin{cases}
reject \hspace{7mm} \textrm{if} \ M \ \textrm{accepts} \ [M]    \\
accept \hspace{7mm}  \textrm{if} \ M \ \textrm{does not accept} \ [M]                     
\end{cases}
\end{equation}
The program-data (i.e., expression-referent) duality is exploited when the machine $M$ runs on the input representing its own description $[M]$. Thus, in this step self-reference is used for the first time, internally. 

\textbf{Step 3.} The final step is to run the inverter (``Liar'') machine $L$ on itself, that is, to consider \textit{self-referential inverter (``Liar'')} $L([L])$:
\begin{equation}
\label{t:dd}
    L([L]) = 
\begin{cases}
reject \hspace{7mm} \textrm{if} \ L \ \textrm{accepts} \ [L]    \\
accept \hspace{7mm}  \textrm{if} \ L \ \textrm{does not accept} \ [L]                     
\end{cases}
\end{equation}
This step uses program-data (i.e., expression-referent) duality and self-reference again, this time externally. It  produces a contradiction in style of the Liar's Paradox: the inverter machine $L$ rejects its input $[L]$ whenever $L$ accepts
$[L]$. This contradiction shows the impossibility of the decider TM $D$, proving undecidability of the Halting problem.

\subsection{Formal systems}

One of the core technical insights of G\"odel was to encode the well-formed formulas (wff's) of a formal system by unique natural numbers. For every wff $W$, this scheme (``G\"odel  numbering'') produces a natural number $\altmathcal{G}(W)$, which is further encoded by a numeral --- the name of the ``G\"odel  number'' of  formula $W$ --- denoted as $\ulcorner \ W \ \urcorner$.

\subsubsection{Gödel's incompleteness theorems}
\label{incomp-1-2}

An important step implicit in  G\"odel's proof of incompleteness theorems is the Self-reference lemma (sometimes referred to as the Fixed-point lemma or the Diagonalisation lemma) ~\cite{Raatikainen,smullyan1984fixed,prokopenko_self-referential_2019}:
\begin{lemma}[\textbf{Self-reference}]
Let $Q(x)$ be an arbitrary formula of formal system $\altmathcal{F}$ with only one free variable. Then there is a sentence (formula without free variables) $W$  such that
\[    \altmathcal{F} \vdash \ W \leftrightarrow Q(\ulcorner \ W \ \urcorner) \ . \]
\end{lemma}
Here, $\altmathcal{F} \vdash V$ denotes that $V$ is a theorem  of the formal system $\altmathcal{F}$, in other words, $V$ is derivable in $\altmathcal{F}$ by a proof. 

According to the Self-reference lemma, for any formula $Q(x)$ that describes a property of a numeral, there exists a sentence $W$ that is logically equivalent to the sentence $Q(\ulcorner \ W \ \urcorner)$. This is self-referential, because the expression $Q(\ulcorner \ W \ \urcorner)$ describes a property of the numeral  $\ulcorner \ W \ \urcorner$ which encodes the G\"odel number of the formula $W$. This also represents a fixed-point, because the formula $W$  is logically equivalent to the formula $Q(\ulcorner \ W \ \urcorner)$ with argument $W$. 

The next formal steps include defining
\begin{itemize}
\item  the formula $\textrm{Proof}_{\altmathcal{F}}(y, x)$ which strongly represents the binary relation ``$y$ is the G\"odel number of a proof of the formula with the G\"odel number $x$'', and 
\item  the provability predicate $\textrm{Provable}_{\altmathcal{F}}(x)$, which captures the property of $x$ being provable in $\altmathcal{F}$, as $\exists y \ \textrm{Proof}_{\altmathcal{F}}(y, x)$.
\end{itemize}

The final step leading to G\"odel's First Incompleteness Theorem applies the Self-reference lemma to the \textit{negated} provability predicate $\neg \textrm{Provable}_{\altmathcal{F}}(x)$: 
\begin{equation}
\label{negProv}
    \altmathcal{F} \vdash \ W \leftrightarrow \neg \textrm{Provable}_{\altmathcal{F}}(\ulcorner \ W \ \urcorner) \ . 
\end{equation}
The obtained expression formally demonstrates that the system $\altmathcal{F}$ can derive that $W$ is true if and only if it is not provable in $\altmathcal{F}$. 

Moreover, an assumption that the system $\altmathcal{F}$ is \textit{consistent} yields that the well-formed formula  $W$ is neither provable nor disprovable in $\altmathcal{F}$, thus demonstrating that the system $\altmathcal{F}$ is \textit{incomplete}~\cite{prokopenko_self-referential_2019}.

Thus, one obtains two incompleteness theorems:
\begin{enumerate}
    \item \textbf{First Incompleteness Theorem}: ``Any consistent formal system $\altmathcal{F}$ within which a certain amount of elementary arithmetic can be carried out is incomplete; i.e., there are statements of the language of $\altmathcal{F}$ which can neither be proved nor disproved in $\altmathcal{F}$.''~\cite{godel_uber_1931}.
    \item \textbf{Second Incompleteness Theorem}: ``For any consistent system $\altmathcal{F}$ within which a certain amount of elementary arithmetic can be carried out, the consistency of $\altmathcal{F}$ cannot be proved in $\altmathcal{F}$ itself.''~\cite{godel_uber_1931}.
\end{enumerate}

\subsubsection{Undecidability}
\label{undeci}
Following \citet{prokopenko_self-referential_2019}, we reproduce here some well-known definitions. A formal system $\altmathcal{F}$ is  \textit{decidable} if the set of its theorems is \textit{strongly representable} in $\altmathcal{F}$ itself: there is some formula $\textrm{P}(x)$ of $\altmathcal{F}$ such that
\begin{equation}
\label{repr}
\begin{aligned}
 \altmathcal{F} \vdash \textrm{P}(\ulcorner \ W \ \urcorner) \ &\mathrm{whenever} \ \altmathcal{F} \vdash W, \ \mathrm{and}\\  
\altmathcal{F} \vdash \neg \textrm{P}(\ulcorner \ W \ \urcorner) \ &\mathrm{whenever} \ \altmathcal{F} \nvdash W \ . 
\end{aligned}
\end{equation}
For \textit{semi-decidability}, i.e., for a \textit{weakly representable} set of theorems, only the first line of (\ref{repr}) is required. 
Nevertheless,  it is possible to construct, within the system $\altmathcal{F}$, a G\"odel sentence $L^{\textrm{P}}$ relative to $\textrm{P}(x)$:
\begin{equation}
\label{phip}
   \altmathcal{F} \vdash \ L^{\textrm{P}} \leftrightarrow \neg \textrm{P}(\ulcorner \ L^{\textrm{P}} \ \urcorner) \ . 
\end{equation}
A contradiction follows --- similar to the contradiction formed by expression (\ref{negProv}). As a result, the strong representability does not hold at least for this sentence, and therefore, $\altmathcal{F}$ must be undecidable.

The well-known G\"odel sentence $L^{\textrm{P}}$ is constructed as $L(\ulcorner \ L(x) \urcorner)$ for some formula $L(x)$ with a free variable. Essentially, this exploits the expression-referent duality within the formal system.  This choice transforms the key expression (\ref{phip}) into
\begin{equation}
\label{vv}
   \altmathcal{F} \vdash \ L(\ulcorner \ L(x) \urcorner) \leftrightarrow \neg \textrm{P}(\ulcorner \ L(\ulcorner \ L(x) \urcorner) \urcorner) \ .
\end{equation}
Thus, self-reference is used both inside and outside of the representative predicate $\textrm{P}(x)$~\cite{Gaifman2007easy}.

\subsection{Dynamical systems: Cellular Automata}
\label{app-CA}

A Cellular Automaton (CA) is a discrete dynamical system defined on an infinite multi-dimensional lattice, so that each lattice site (cell) takes a value from a finite alphabet~\cite{Wolfram1984,goldenfeld}. Each cell is updated in discrete time steps according to a deterministic local rule involving values of neighbouring cells. A configuration of cells in the lattice is an infinite sequence of specific cell values at a given time, being completely determined by the preceding configuration, while the initial configuration is a sequence of lattice cells at time zero. 

A repeated application of a update rules, starting from the initial configuration, produces an evolution of configurations over time. It is important to restrict the space of possible configurations to certain subspaces, e.g., recursive configurations, where a cell state is assigned by a computable function. In this case, the restriction produces an ``effective dynamical system''~\cite{Sutner-chapter}.

\citet{prokopenko_self-referential_2019} extended the definition of a CA to include a termination condition, by adopting a convention which determines when a specific (desired) output has occurred. This corresponds to tracing the configuration dynamics, starting from the initial configuration, until a (primitive recursively decidable) ``halting'' condition is applicable, e.g., an attractor (a fixed-point or a limit cycle) is formed~\cite{Lindgren1990,Sutner-chapter}.  CA $M$ starting from some initial configuration $I$ and provided with a termination condition $m$ can be interpreted as an automaton performing computation $M: I \rightarrow m$, rather than simply processing information.

\subsubsection{Universal Cellular Automata}
\label{UCA}

A universal CA (UCA) is a CA which can simulate dynamics of any CA. In order to construct a UCA we must encode any simulated CA (including its termination condition and the initial configuration), as data that can be represented in the initial configuration for the UCA. 
An important step used in simulating CAs is the \textit{coarse-graining} of the CA dynamics, by grouping neighbouring cells into a supercell of specific dimensions~\cite{goldenfeld}. Supercells exhibit a fixed number of distinct higher-order patterns, essentially forming a new, higher-order, alphabet~\cite{wiki:otca}. 

\begin{figure}[!h]
    \centering
\includegraphics[width=1.0\columnwidth]{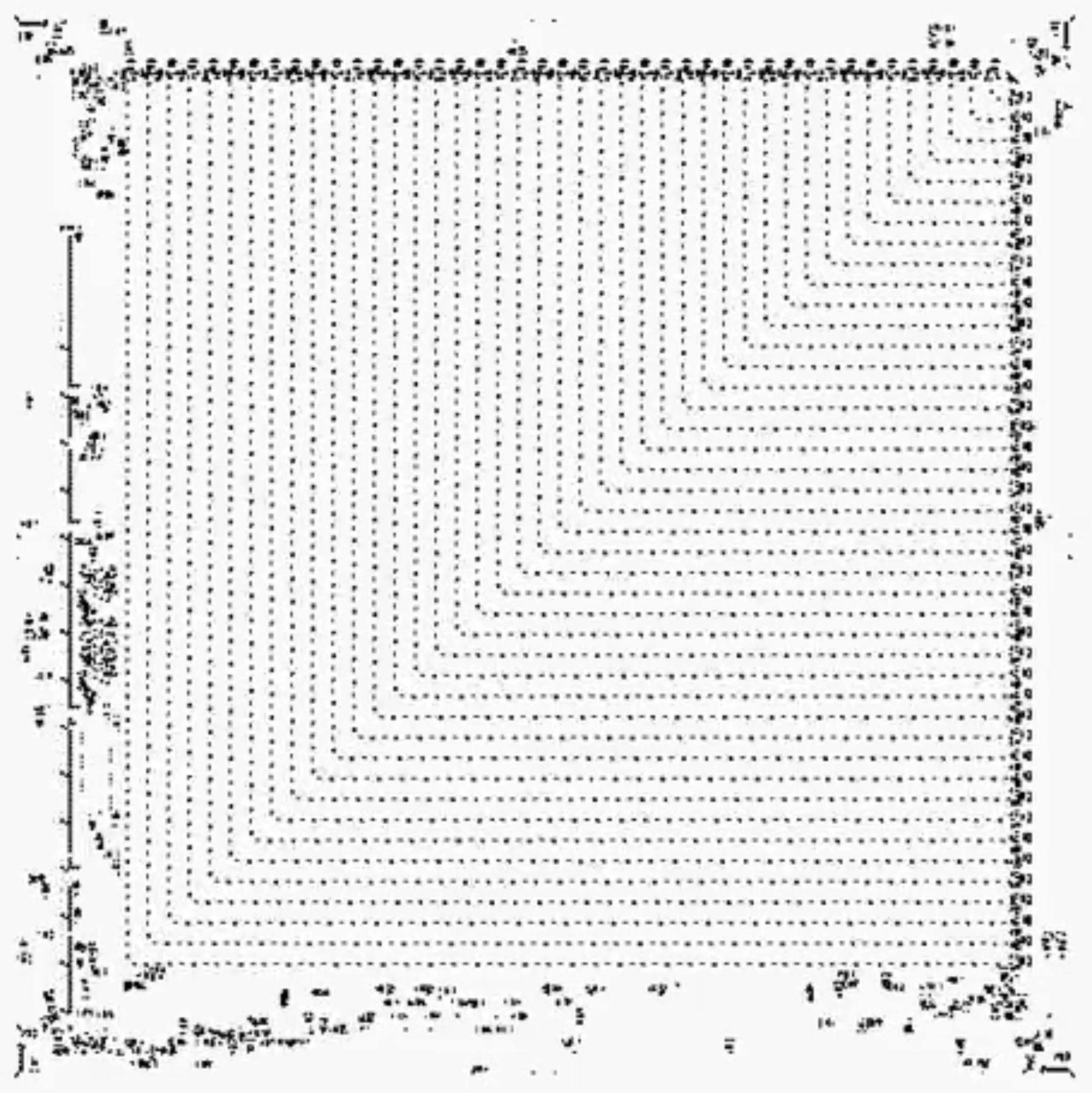}
    \caption{The OTCA metapixel is a $2048 \times 2048$ unit cell which can emulate any Life-like cellular automata, shown in ON state~\cite{wiki-OTCA}.}
    \label{fig:otcametapixel}
\end{figure}

\begin{figure}[!h]
    \centering
\includegraphics[width=1.0\columnwidth]{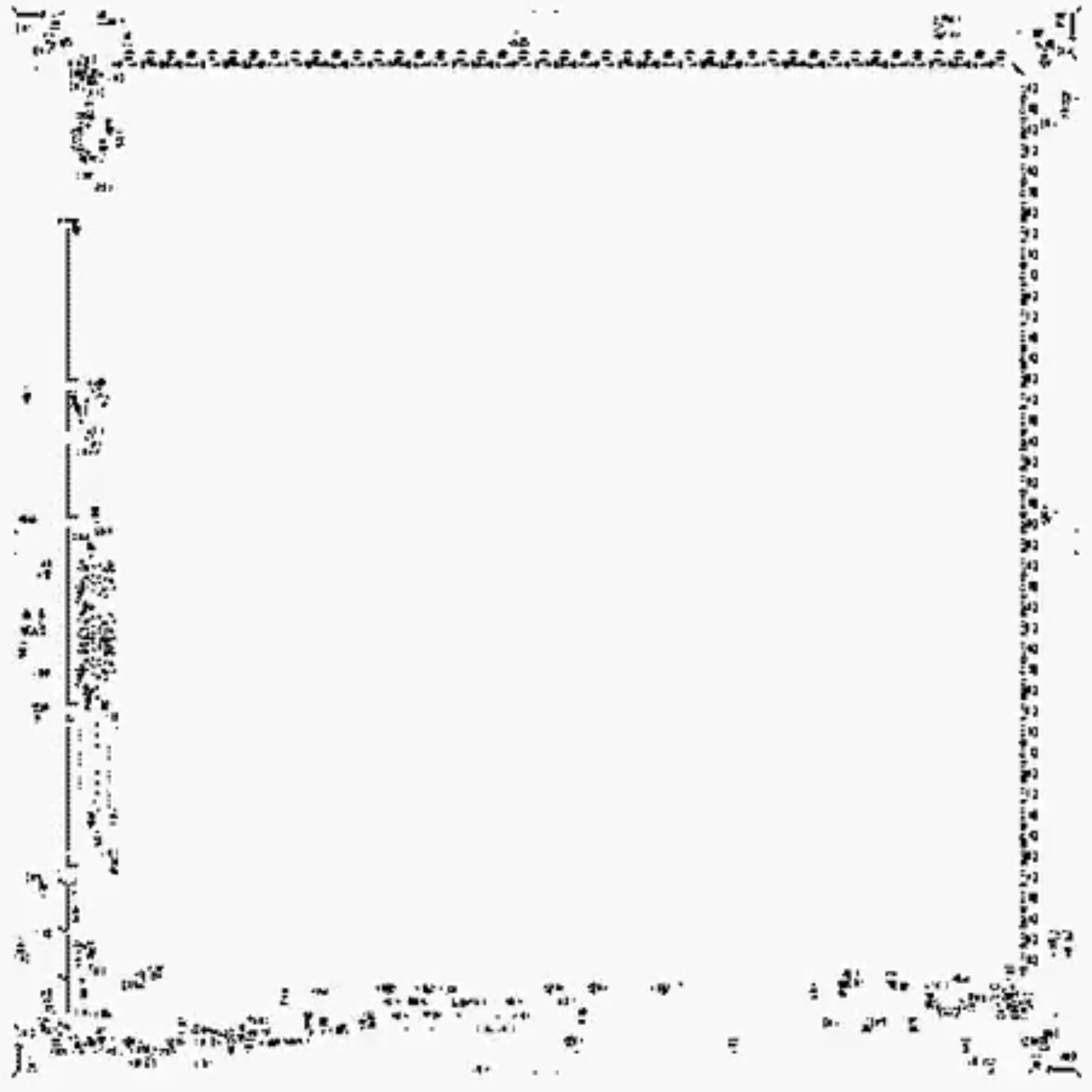}
    \caption{The OTCA metapixel shown in OFF state~\cite{wiki-OTCA}.}
    \label{fig:otcametapixel-off}
\end{figure}

An impressive supercell example --- the \emph{Outer Totalistic Cellular Automata metapixel} (\emph{OTCA metapixel}) --- was designed to simulate a canonical two-dimensional CA ``Game of Life''~\cite{Gardner1970}, within the ``Life in Life'' simulation~\cite{wiki:otca}. Figures~\ref{fig:otcametapixel} and \ref{fig:otcametapixel-off} illustrate the emulated ON and OFF states of the OTCA metapixel, which alternate after 35,328 generations. 

\begin{figure}[!h]
    \centering
\includegraphics[width=1.0\columnwidth]{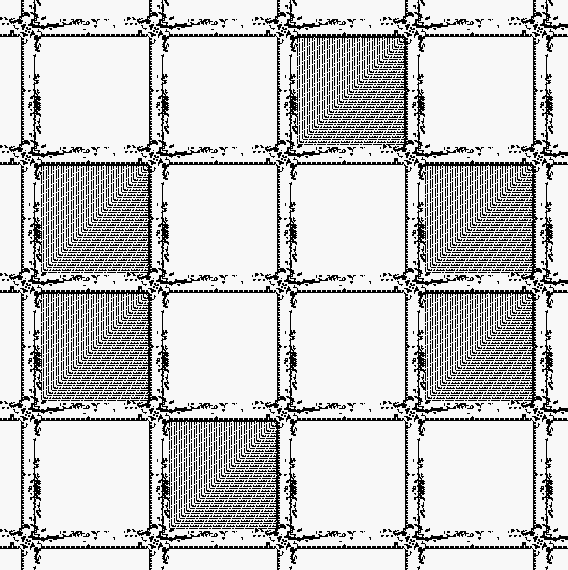}
    \caption{A coarse-grained pattern known in Conway's Game of Life as ``toad'' (oscillating polyomino), emulated by metapixel cells~\cite{Greene2006}.}
    \label{fig:meta-toad}
\end{figure}

Coarse-grained supercells (OTCA metapixels) appear to interact in accordance with the rules of the ``Game of Life'' (for example, see Fig.~\ref{fig:meta-toad}). However, following \citet{prokopenko_self-referential_2019}, we would like to point out that the coarse-grained metapixel states and coarse-grained patterns emerge as a result of the dynamics generated by the underlying fine-grained CA, rather than by any direct interaction between metapixels themselves --- in other words, coarse-grained dynamics is entirely simulated. 

\subsubsection{Undecidable Cellular Automata}
\label{undecidable-CA}

Revisiting the ``Life in Life'' simulation~\cite{wiki:otca}, we note that any  termination condition specified for the fine-grained ``Game of Life'' (e.g., reaching a particular attractor or a predefined configuration) may also be defined at the coarse-grained level, in terms of the metapixel states reaching an identical (but coarse-grained) attractor or configuration. In other words, the termination condition for the ``Life in Life'' simulation is specified in the coarse-grained alphabet. 

Using a UCA such as the one based on the OTCA metapixel, it is possible to repeat the steps described in Section~\ref{inverter_liar}, and produce a self-referential Liar CA. In the ``Life in Life'' example, the self-referential coarse-grained Liar would \textit{accept} those attractors or configurations that the fine-grained ``Game of Life'' was defined to \textit{reject}, and vice versa~\cite{prokopenko_self-referential_2019}. 

In other words, the Liar paradox is formed in the coarse-grained CA dynamics by (1) inverting (flipping) the termination conditions, thus, utilising \textit{negation}; and (2) setting the coarse-grained Liar to run on the initial configuration that encodes itself (that is, simulating \textit{itself}), thus, utilising the \textit{expression-referent} or \textit{dynamics-data duality}.  

Several universal CAs have been constructed, e.g., the elementary CA rule $110$~\cite{Cook2004} and ``Life in Life''~\cite{wiki:otca}. Universal deciders are, however, impossible. Thus, undecidable dynamics, which corresponds to \textit{strange attractors} and \textit{the edge of chaos}~\cite{casti91}, means that it is impossible to specify termination conditions which would assign binary outcomes for any conceivable CA $M$ simulated from an initial configuration $I$~\cite{prokopenko_self-referential_2019}.

\section{Equivalences across frameworks}

The comparative analysis of factors contributing to computational undecidability across Turing machines, formal systems and cellular automata~\cite{prokopenko_self-referential_2019} is summarised in Table~\ref{Summary-table}.

\onecolumn
\begin{table}[h]
\begin{center}
\renewcommand{\arraystretch}{1.5}
\begin{tabular}{|p{5.6cm}|p{5.6cm}|p{5.6cm}|}
\hline
 \textbf{Formal systems} & \textbf{Turing machines} & \textbf{Cellular Automata}  \\ 
\hline 
\hline
\multicolumn{3}{|c|}{\textit{State-space}}\\
\hline
Alphabet & Input and tape alphabets  &  Alphabet \\
\hline
Symbol strings & Tape strings  & Configurations in state-space \\
\hline
Grammar & Admissible syntax & Constraints on state-space \\
\hline
Well-formed formula (wff), \newline restricted by grammar & Recognisable tape pattern & Primitive recursively decidable configuration  \\
\hline
Infinite language & Infinite tape & Infinite lattice \\ 
\hline
\hline
\multicolumn{3}{|c|}{\textit{Problem definition and inferences/computation/dynamics}}\\
\hline
Axioms & (A part of) initial tape & (A part of) initial configuration \\
\hline
($\ddagger$) Target: wff to be proven & ($\ddagger$) Target:  string as part of initial tape & ($\ddagger$) Target: subset of initial configuration \\
\hline
($\ddagger$) Proving or disproving  target wff & ($\ddagger$) Final states $accept$ and $reject$ & ($\ddagger$) Termination condition testing against predefined configurations \\
\hline
Rules of inference & Transition function &  Local update rules  \\
\hline
Proof: derivation sequence & Sequence of tape patterns and machine states & Dynamic evolution of configurations \\
\hline
($\dagger$) An external criterion distinguishing a wff in a proof & ($\dagger$) Final state $halt$ & ($\dagger$) Termination condition testing for fixed points or limit cycles \\
\hline
($\dagger$) Theorem: the last wff in a proof & ($\dagger$) Final output written on the tape & ($\dagger$) The attractor configuration(s) \\
\hline
\hline
\multicolumn{3}{|c|}{\textit{Proof of undecidability}}\\
\hline
Weakly representative predicate  & Universal TM & Universal CA \\
\hline 
($\nexists$)  Representative predicate $\textrm{P}(x)$ & ($\nexists$)  Universal decider TM $D$ & ($\nexists$)  Universal decider CA $D$ \\
\hline
$\altmathcal{G}(W(x))$: G\"odel number of $W(x)$  & Encoding of TM $M$, denoted $[M]$ & Encoding of CA $M$, denoted  $[M]$ \\
\hline
Unique decoding of $W(x)$ from G\"odel number $\altmathcal{G}(W(x))$ & Unique decoding of TM $M$ from $[M]$ & Unique decoding of CA $M$ from $[M]$ \\
\hline
First (internal) self-referencing: $W(\ulcorner \ W(x) \ \urcorner)$ & First (internal)  self-referencing: $M[M]$ &  First (internal)  self-referencing: $M[M]$ \\
\hline
Diagonalisation term for $W(x)$: $\textrm{diag}(\ulcorner  W(x)  \urcorner) = \ulcorner  W(\ulcorner  W(x)  \urcorner)  \urcorner$ 
& Compound encoding of TM $M$, as $[M[M]]$ & Compound encoding of CA $M$, as $[M[M]]$ \\ 
\hline
Inverted (``Liar'') predicate \newline $L^{\textrm{P}}(x) \equiv \neg \textrm{P}_{\altmathcal{F}}(\textrm{diag}(x) )$ & Inverter (``Liar'') TM $L([M])$ & Inverter (``Liar'') CA $L$ running on $[M]$ \\ 
\hline
G\"odel sentence \newline $L^{\textrm{P}}(x) = L(\ulcorner \ L(x) \urcorner)$ & Self-referential inverter (``Liar'') TM $L([L])$  & Self-referential inverter (``Liar'') CA $L$ running on $[L]$ \\
\hline
Second (external) self-referencing: \newline $\altmathcal{F} \sststile{}{?}  \ L^{\textrm{P}} \leftrightarrow \neg \textrm{P}(\ulcorner \ L^{\textrm{P}} \ \urcorner)$ & Second (external)  self-referencing: \newline $L([L]) = \ ?$ & Second (external)  self-referencing: \newline $L: [L] \rightarrow \ ?$ \\ 
\hline
\hline
G\"odel Incompleteness Theorem, leading to undecidability & The Halting problem & Undecidable dynamics and the ``edge of chaos'' \\
\hline
\end{tabular}
\end{center}
\caption{\label{Summary-table} Comparison of state-spaces, problem definitions and proofs of undecidability across three computational frameworks. Adapted from \citet{prokopenko_self-referential_2019}.}
\end{table}
\twocolumn


\end{document}